\documentclass[trackchanges,twocolumn]{aastex701}

\usepackage{graphicx}
\usepackage{soul}
\hypersetup{
    colorlinks=true,
    linkcolor=blue,
    citecolor = blue,
    filecolor=magenta,
    urlcolor=blue}
\usepackage{booktabs}
\usepackage{subcaption} 
\usepackage{rotating}
\usepackage{xcolor}
\usepackage{cancel}
\begin{document}

\title{CRIRES+ reveals  the chemistry of the stellar sub-populations in the bulge fossil fragment Liller 1\footnote{Based on observations collected at the Very Large Telescope of the European Southern Observatory at Cerro Paranal (Chile) under program 109.230K and Large Program 110.24A4 (PI:Ferraro)}}

\author[orcid=0009-0008-8156-724X]{L. Chiappino}
\affiliation{Dipartimento di Fisica e Astronomia, Università degli Studi di Bologna, Via Gobetti 93/2, I-40129 Bologna, Italy}
\affiliation{INAF, Osservatorio di Astrofisica e Scienza dello Spazio di Bologna, Via Gobetti 93/3, I-40129 Bologna, Italy}
\email[show]{lan.chiappino@unibo.it}

\author[orcid=0000-0002-6040-5849]{L. Origlia} 
\affiliation{INAF, Osservatorio di Astrofisica e Scienza dello Spazio di Bologna, Via Gobetti 93/3, I-40129 Bologna, Italy}
\email{livia.origlia@inaf.it}

\author[orcid=0000-0002-4639-1364]{C. Fanelli} 
\affiliation{INAF, Osservatorio di Astrofisica e Scienza dello Spazio di Bologna, Via Gobetti 93/3, I-40129 Bologna, Italy}
\email{cristiano.fanelli@inaf.it}

\author[orcid=0009-0000-6641-1695]{A. Bartolomei}
\affiliation{Dipartimento di Fisica e Astronomia, Università degli Studi di Bologna, Via Gobetti 93/2, I-40129 Bologna, Italy}
\affiliation{INAF, Osservatorio di Astrofisica e Scienza dello Spazio di Bologna, Via Gobetti 93/3, I-40129 Bologna, Italy}
\email{alessia.bartolomei4@unibo.it} 

\author[orcid=0000-0002-2165-8528]{F.R.Ferraro}
\affiliation{Dipartimento di Fisica e Astronomia, Università degli Studi di Bologna, Via Gobetti 93/2, I-40129 Bologna, Italy}
\affiliation{INAF, Osservatorio di Astrofisica e Scienza dello Spazio di Bologna, Via Gobetti 93/3, I-40129 Bologna, Italy}
\email{francesco.ferraro3@unibo.it} 

\author[orcid=0000-0001-5613-4938]{B. Lanzoni}
\affiliation{Dipartimento di Fisica e Astronomia, Università degli Studi di Bologna, Via Gobetti 93/2, I-40129 Bologna, Italy}
\affiliation{INAF, Osservatorio di Astrofisica e Scienza dello Spazio di Bologna, Via Gobetti 93/3, I-40129 Bologna, Italy}
\email{barbara.lanzoni3@unibo.it} 

\author[orcid=0000-0002-7104-2107]{C. Pallanca}
\affiliation{Dipartimento di Fisica e Astronomia, Università degli Studi di Bologna, Via Gobetti 93/2, I-40129 Bologna, Italy}
\affiliation{INAF, Osservatorio di Astrofisica e Scienza dello Spazio di Bologna, Via Gobetti 93/3, I-40129 Bologna, Italy}
\email{cristina.pallanca3@unibo.it} 

\author[orcid=0000-0002-5038-3914]{M. Cadelano}
\affiliation{Dipartimento di Fisica e Astronomia, Università degli Studi di Bologna, Via Gobetti 93/2, I-40129 Bologna, Italy}
\affiliation{INAF, Osservatorio di Astrofisica e Scienza dello Spazio di Bologna, Via Gobetti 93/3, I-40129 Bologna, Italy}
\email{mario.cadelano@unibo.it} 

\author[orcid=0000-0002-0845-6171]{D. Romano} 
\affiliation{INAF, Osservatorio di Astrofisica e Scienza dello Spazio di Bologna, Via Gobetti 93/3, I-40129 Bologna, Italy}
\email{donatella.romano@inaf.it}

\author[orcid=0000-0003-4237-4601]{E. Dalessandro} 
\affiliation{INAF, Osservatorio di Astrofisica e Scienza dello Spazio di Bologna, Via Gobetti 93/3, I-40129 Bologna, Italy}
\email{emanuele.dalessandro@inaf.it}

\author[orcid=0000-0001-8892-4301]{D. Massari} 
\affiliation{INAF, Osservatorio di Astrofisica e Scienza dello Spazio di Bologna, Via Gobetti 93/3, I-40129 Bologna, Italy}
\email{davide.massari@inaf.it}

\author[orcid=0000-0002-6092-7145]{E. Valenti}
\affiliation{European Southern Observatory, Karl-Schwarzschild-Strasse 2, 85748 Garching bei Munchen, Germany}
\affiliation{Excellence Cluster ORIGINS, Boltzmann-Strasse 2, D-85748 Garching Bei Munchen, Germany}
\email{evalenti@eso.org} 

\author[orcid=0000-0003-0427-8387]{R.M. Rich}
\affiliation{Department of Physics and Astronomy, UCLA, 430 Portola Plaza, Box 951547, Los Angeles, CA 90095-1547, USA}
\email{rmrastro@gmail.com} 

\begin{abstract}
In this paper we present the chemical screening of the complex stellar population discovered in the Bulge Fossil Fragment Liller~1. This study is part of the {\it Bulge Cluster Origin (BulCO)} survey based on a Large Program at the ESO-VLT with the high resolution spectrograph CRIRES+. The survey is aimed at performing an unprecedented chemical screening of 17 stellar systems orbiting the Milky Way bulge, with the ultimate goal of unveiling their origin and true nature.  We measured precise chemical abundances of iron, CNO, iron-peak,  $\alpha$- other light-elements, and neutron-capture elements for a sample of 30 red giant branch stars, kinematic members of Liller~1. 
The presented analysis provides the high-resolution spectroscopic proof of the complex chemistry of this massive stellar system, with multi-metallicity sub-populations of different ages that nicely fits into a self-enrichment scenario. We find no evidence for the Na-O anticorrelation associated with genuine globular clusters; rather the overall abundance trends are similar to those seen in the bulge field and in Terzan 5, providing definitive evidence of an in-situ formation of Liller 1 within the Galactic bulge.
\end{abstract}


\section{Introduction} 
\label{intro}
The formation of galaxy bulges is a topic largely debated in the literature \citep[see, e.g.,][]{immeli_04, dek09,ger12,saha13,kalita22,tan24}. Among the proposed models, one fascinating possibility is that  bulges form through the merging of primordial clumps of stars and gas \citep[see, e.g.,][]{immeli_04, elme08, bournaud09, bournaud16}. 
Numerical simulations show that such massive clumps (with masses of 10$^8$-10$^9$ M$_{\odot}$) can form from  disk instabilities in gas-rich galaxies, and/or by the clustering of smaller, seed clumps of  10$^7$-10$^8$ M$_{\odot}$ \citep[e.g.,][]{beh16}. 
This scenario is observationally supported by the so-called chain and clumpy galaxies discovered at high-redshift \citep{elme09, genzel11, tacchella15}, that actually show giant clumps of stars and gas on the verge of merging.   
The vast majority of these primordial structures are destined to merge over short timescales (a few 10$^8$ yr) to form the galaxy spheroid. However, simulations \citep{bournaud16} also show  that some fragments 
of those pristine massive clumps survive the total disruption and evolve as independent stellar systems.   These surviving structures might still be present in the inner regions of the host galaxy under the appearance of massive GCs, but hosting multi-iron and multi-age subpopulations sharing the same chemical patterns of the bulge field stars.  The identification and the characterization of those systems is of paramount importance for our understanding of the assembling process of the galaxy.

In this context, we performed an unprecedented photometric and spectroscopic investigation of Terzan 5 and Liller 1, two stellar systems traditionally cataloged as bulge GCs, and we discovered that they host, instead, sub-populations with age differences of a few Gyr and iron abundances spanning almost 1 dex (see \citealp{Ferraro_09, ferraro_16, Origlia_11, Origlia_13, origlia_19, Massari_14, origlia+25} for Terzan 5, and \citealp{Ferraro_09, ferraro_21, pallanca_21, crociati_23, deimer24,fanelli+24,ferraro+25a} for Liller 1). Their reconstructed star formation histories are characterized by a continuous low-intensity star formation activity with multiple bursts (see \citealt{dalessandro_22, crociati+24,zullo+26}), and their observed chemical abundances well fit into a self-enrichment scenario \citep{romano23}. Their [$\alpha$/Fe]-[Fe/H] patterns closely follow that traced by bulge field stars that, in turn, is radically different from those observed in  the Milky Way halo and disk, and in any dwarf galaxy of the Local Group (see \citealp{origlia+25, ferraro+25a}, and references therein). This clearly demonstrates that neither Terzan 5, nor Liller 1 are genuine GCs, and strongly suggests that they are "bulge fossil fragments" (BFFs, \citealt{ferraro_21}), i.e., the fossil remains of two massive primordial clumps that contributed to the early bulge formation. The importance of understanding the formation history of these peculiar stellar systems has been recently amplified by the suggestion that they could represent the most efficient factories of gravitational waves in the Galaxy \citep{ferraro+26}.
 
A recent chemical study of Liller~1 based on APOGEE spectra for 14 very bright giant stars, half of which being located in the outskirts of the system, claims significant differences with respect to the composition of bulge stars, and therefore suggests a possible extra-Galactic origin for this object \citep{liptrott26}.
Here, we present the results of a detailed, high-resolution chemical screening of the complex stellar populations discovered in 
Liller 1, 
performed with the high-resolution spectrograph CRIRES+ \citep{kaufl+04, dorn+14, dorn+23} mounted at the ESO Very Large Telescope.
Sect.~\ref{obs} describes the observations and data reduction, as well as the properties of the selected targets. Sect.~\ref{spec} presents  the spectral analysis that has been performed and the obtained radial velocities, stellar parameters and chemical abundances for the observed stars, while in Sect.\ref{disc} we discuss the results and draw our conclusions about Liller~1 formation and evolution scenarios.

\begin{figure*}
    \centering
    \includegraphics[width=\textwidth]{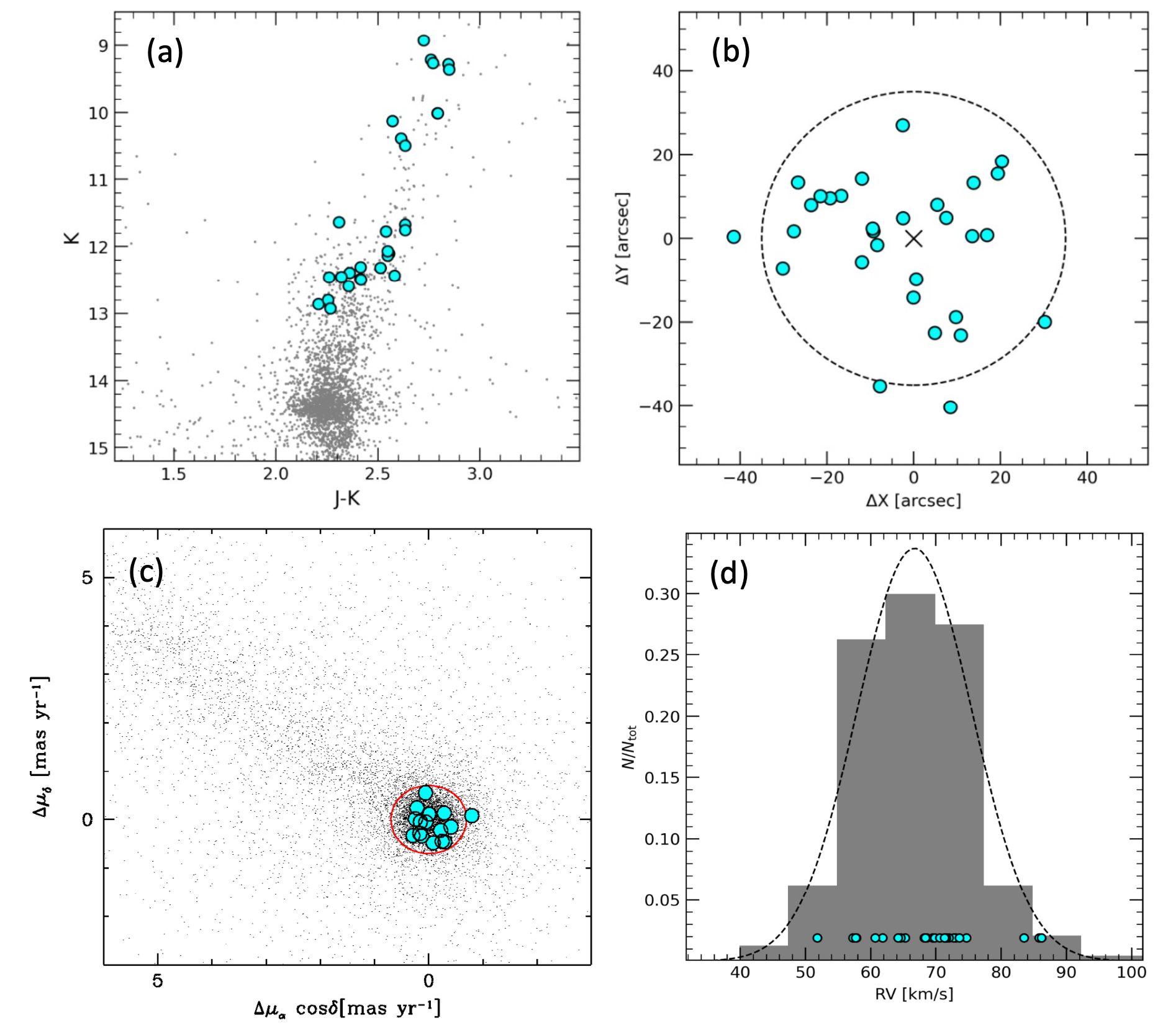}
    \caption{Properties of the Liller~1 spectroscopic targets for which we measured chemical abundances in this work (cyan circles in all the panels).
    Panel (a): position of the targets in the near-infrared CMD of Liller 1 (gray dots; from \citealp{valenti10} and \citealp{saracino_15}). Panel (b): position of the targets in the plane of the sky with respect to the cluster center (marked with a cross). The dashed circle has a radius equal to the half-mass radius $r_h=30.5"$ (\citealt{saracino_15}). Panel (c): vector-point diagram of relative proper motions toward Liller~1 computed from the combination of HST and ground-based adaptive optics data. The grey dots correspond to stars brighter than $K=16$ in the sampled field of view. The red circle encloses the stars classified as likely Liller 1 members in \citet{dalessandro_22}. Panel (d): radial velocity distribution of the 30 targets, compared to the overall distribution of member stars (grey histogram and dashed Gaussian fitting) from previous studies \citep{crociati_23,deimer24,fanelli+24}.}
    \label{fig1}
\end{figure*}

\section{Observations and data reduction}
\label{obs}
We acquired high resolution CRIRES+ spectra of 30 giant stars in Liller~1 between June 2022 and July 2024.
Fourteen targets were observed in the context of the {\it Bulge Cluster Origin} (BulCO) survey \citep[see][]{ferraro+25a} under the Large Program 110.24A4 (PI: Ferraro), using two different setups: gratings J1226 (1116 $-$ 1356 nm) and H1582 (1484 $-$ 1854 nm) for the four brightest stars; gratings K2166 (1921 $-$ 2472 nm) and H1582 (1484 $-$ 1854 nm) for the other ten stars. Sixteen targets were observed with grating K2166 (1921 $-$ 2472 nm) under program 109.230K (PI:Ferraro), a pilot project for the BulCO survey.
These gratings allow the sampling of several unblended atomic spectral lines of iron, iron-peak,  $\alpha$-elements, a few other light and neutron-capture elements. In addition, a few dozen CO, OH and CN molecular lines are sampled, allowing the measure of C, N, O abundances.
In all cases we used the 0.4$\arcsec$ slit, which provided us with an overall spectral resolution of $R\sim 50,000$. For the data reduction, we used the CR2RES version 1.4.1 pipeline,\footnote{\url{https://www.eso.org/sci/software/pipelines/cr2res/cr2res-pipe-recipes.html}} which performs dark and flat-field corrections, sky-subtraction using nod pairs, wavelength calibration through arc lamps, and finally adopts the optimum extraction method to obtain 1D spectra. The signal-to-noise ratio per resolution element of the final extracted spectra is always $\ge$40. 

\subsection{Properties of the target sample}
The coordinates, $K$-band magnitudes (from \citealp{valenti10}, \citealp{saracino_15}, and \citealp{ferraro_21}), atmospheric parameters and radial velocities (obtained as described below) for the 30 observed targets are listed in Table \ref{tab1}. 
The sample also includes the 12 stars for which Ca, Mg, Si and Fe abundances were already discussed in \citet{ferraro+25a}.
The top-left panel of Fig.~\ref{fig1} shows the positions of the 30 stars in the near-infrared color-magnitude diagram (CMD).
They span the red giant branch (RGB) from the tip down to about 1.5 mag above the red clump, thus probing a rather wide range of stellar parameters.
The ten brightest stars are in common with the sample of \citet{deimer24} and four of these have been also investigated in \citet{fanelli+24}.
The spatial distribution of the selected stars with respect to the system center (RA=$17^{\rm h}$ $33^{\rm m}$ $24.56^{\rm s}$, Dec=$-33^\circ$ $23\arcmin$ $22.40\arcsec$; \citealp{saracino_15}) is plotted in the top-right panel of Fig. \ref{fig1}, showing that all targets are located in the  central region of Liller~1 ($r<40\arcsec$), reaching distances slightly larger than the system half-mass radius ($r_h=30.5\arcsec$; see \citealp{saracino_15}).

\begin{table}[t]
\centering
\caption{Coordinates, $K$-band magnitude, atmospheric parameters, and radial velocity of the observed stars in Liller~1.}
\label{tab1}
\scriptsize
\setlength{\tabcolsep}{3pt}
\renewcommand{\arraystretch}{1.1}
    \begin{tabular}{|c|c|c|c|c|c|c|}
    \hline\hline
    ID &  RA   &  Dec   &   K & T$_{\rm eff}$   & log(g) & RV \\
    \hline
       & [Deg] & [Deg]  & [mag]  & K  & dex & km/s  \\
    \hline
        100157 &   263.3536852 &  -33.3958297  & 10.01 &   3550 &       0.50 &      71.5 \\
        100437 &   263.3524868 &  -33.3922546  & 11.64 &   4150 &       1.50 &      73.0\\
        100571 &   263.3546688 &  -33.4007517  & 11.75 &   4000 &       1.25 &      68.4 \\
        100658 &   263.3607202 &  -33.3951039  & 12.07 &   4100 &       1.25 &      57.8 \\
        100689 &   263.3553501 &  -33.3959916  & 12.13 &   4100 &       1.50 &      72.7 \\
        100756 &   263.3523060 &  -33.3934947  & 12.39 &   4150 &       1.50 &      72.0 \\
        100760 &   263.3501594 &  -33.3993386  & 12.49 &   4200 &       1.25 &      68.6 \\
        100901 &   263.3499944 &  -33.3899851  & 12.59 &   4200 &       1.75 &      57.7 \\
        100987 &   263.3550231 &  -33.3947850  & 12.43 &   4150 &       1.50 &      61.8 \\
        200119 &   263.3439575 &  -33.3915306  & 10.38 &   3700 &       0.75 &      68.2 \\
        200179 &   263.3490171 &  -33.3911168  & 10.01 &   3550 &       0.75 &      69.5 \\
        300094 &   263.3408120 &  -33.3894537  & 9.36 &   3400 &       0.25 &      60.7 \\
        300097 &   263.3490175 &  -33.3856006  & 10.13 &   3600 &       0.50 &      69.8 \\
        300162 &   263.3476816 &  -33.3867406  & 10.49 &   3700 &       0.75 &      57.3 \\
        300315 &   263.3463589 &  -33.3867729  & 9.22 &   3350 &       0.25 &      70.6 \\
        300553 &   263.3497432 &  -33.3890865  & 11.77 &   4050 &       1.25 &      68.4 \\
        300614 &   263.3457652 &  -33.3873565  & 12.45 &   4150 &       1.75 &      74.7 \\
        300682 &   263.3497083 &  -33.3888985  & 12.46 &   4150 &       1.50 &      73.6 \\
        300701 &   263.3469751 &  -33.3868982  & 12.32 &   4150 &       1.50 &      57.7 \\
        300727 &   263.3449122 &  -33.3858666  & 12.50 &   4200 &       1.75 &      71.6 \\
        387099 &   263.3446613 &  -33.3890830  & 8.92 &   3400 &       0.50 &      71.3 \\
        400065 &   263.3579614 &  -33.3844441  & 9.26 &   3400 &       0.50 &      85.8 \\
        400087 &   263.3561627 &  -33.3858685  & 9.28 &   3400 &       0.50 &      86.2 \\
        400476 &   263.3516267 &  -33.3820491  & 11.77 &   4000 &       1.25 &     51.8 \\
        400519 &   263.3577128 &  -33.3852708  & 11.67 &   4000 &       1.25 &     65.3 \\
        400733 &   263.3560726 &  -33.3894022  & 12.31 &   4150 &       1.50 &     64.6 \\
        400778 &   263.3544150 &  -33.3881915  & 12.86 &   4300 &       1.75 &      71.6 \\
        400829 &   263.3516455 &  -33.3882126  & 12.80 &   4250 &       1.75 &      64.2 \\
        400860 &   263.3570144 &  -33.3893289  & 12.10 &   4100 &       1.25 &     83.5 \\
        400887 &   263.3538377 &  -33.3873272  & 12.93 &   4300 &       2.00 &      71.3 \\
\hline\hline
\end{tabular}
\end{table}

Seventeen out of 20 stars with $K>11$ have proper motions measured from a combination of Hubble Space Telescope (HST) and Gemini Multi-Conjugate Adaptive Optics data \citep{ferraro_21, dalessandro_22}. The bottom-left panel of  Fig.~\ref{fig1} shows their location in the vector-point diagram compared to that of stars brighter than $K=16$ in the sampled field of view. They are all included with a circle of radius 0.7 mas yr$^{-1}$ centered on (0,0), which corresponds to the 1 $\sigma$ dispersion around the bulk motion, and thus selects likely members to Liller 1 (see \citealp{dalessandro_22}), while Galactic field stars show an elongated proper motion distribution that extends to much larger values in both the components.
Among the 13 stars with no high-resolution proper motion measurement, three have one epoch only, while the brightest ten are heavily saturated in the HST images.

\begin{table}[t]
\caption{Portion of the line list used for chemical analysis, including identification, wavelength ($\lambda$, in \AA), excitation potential (EP, in eV), and oscillator strength ($\log gf$) of each transition. The full table is available in machine-readable form in the {\it Astrophysical Journal} online edition.}
\label{tab2}
\scriptsize
\setlength{\tabcolsep}{3pt}
\renewcommand{\arraystretch}{0.9}
\begin{tabular}{|c|c|c|c|}
\hline\hline
Transition ID & $\lambda$ & EP & $\log gf$ \\
\hline
        & [\AA]     & [eV] & [dex]       \\
\hline
Fe I & 14988.778 & 6.169 & 0.186 \\
Fe I & 15013.771 & 6.222 & 0.087 \\
Fe I & 15239.712 & 6.419 & -0.032 \\
$\cdots$ & $\cdots$ & $\cdots$ & $\cdots$ \\
\hline\hline
\end{tabular}
\vspace{0.00cm} 
\end{table}

However, the latter have Gaia proper motions and their likely membership to Liller~1 has been already discussed in \citet{deimer24} and \citet{fanelli+24}. Thus, only three stars in our CRIRES+ sample (namely, 100437, 100901, and 300553) have no proper motion measurements.
However, their location in the central region of the system, where member stars largely dominate over field interlopers (see \citealp{ferraro_21}), together with their positions along the RGB and their measured radial velocities (see Table \ref{tab1}), strongly support their classification as likely members of Liller~1.

\section{Spectral analysis}
\label{spec}
Spectral analysis was performed via cross-correlation and spectral synthesis techniques.
The synthetic spectra used to derive the radial velocities, stellar parameters and chemical abundances of the target stars were computed using the radiative transfer code TURBOSPECTRUM \citep{1998A&A...330.1109A,2012ascl.soft05004P} under the LTE approximation, the atomic line list from the VALD3 database \citep{2015BaltA..24..453R} that also includes line hyperfine structuring for odd elements, molecular line data from B. Plez’s online compilation,\footnote{\url{https://www.lupm.in2p3.fr/users/plez/}} and MARCS model atmospheres \citep{2008A&A...486..951G}.

We computed grids of synthetic spectra in the $J$, $H$, and $K$ bands, with effective temperatures T$_{\rm eff}$ ranging from 3000 K to 5000 K in steps of 50 K,  surface gravities $\log(g)$ from 0.00 to 3.50 in steps of 0.25, and metallicities [Fe/H] from $-1.0$ to +0.50 dex in steps of 0.25 dex. We considered both solar-scaled and $\alpha$-enhanced compositions, also including different levels of nitrogen enhancement and carbon depletion. 
A microturbulent velocity of 2 $\pm$ 0.3 km s$^{-1}$ was assumed for all the 30 targets, as this value is characteristic of bulge giant stars with comparable effective temperatures and metallicities. Subsequently, the synthetic spectra were convolved with a Gaussian profile corresponding to the line broadening measured for each individual observed star, with
a FWHM that varied between 9 and 12 km s$^{-1}$ due to macroturbulence, for an optimal match with the observed spectra.

\begin{figure}[]
    \centering
    \includegraphics[width=\columnwidth]{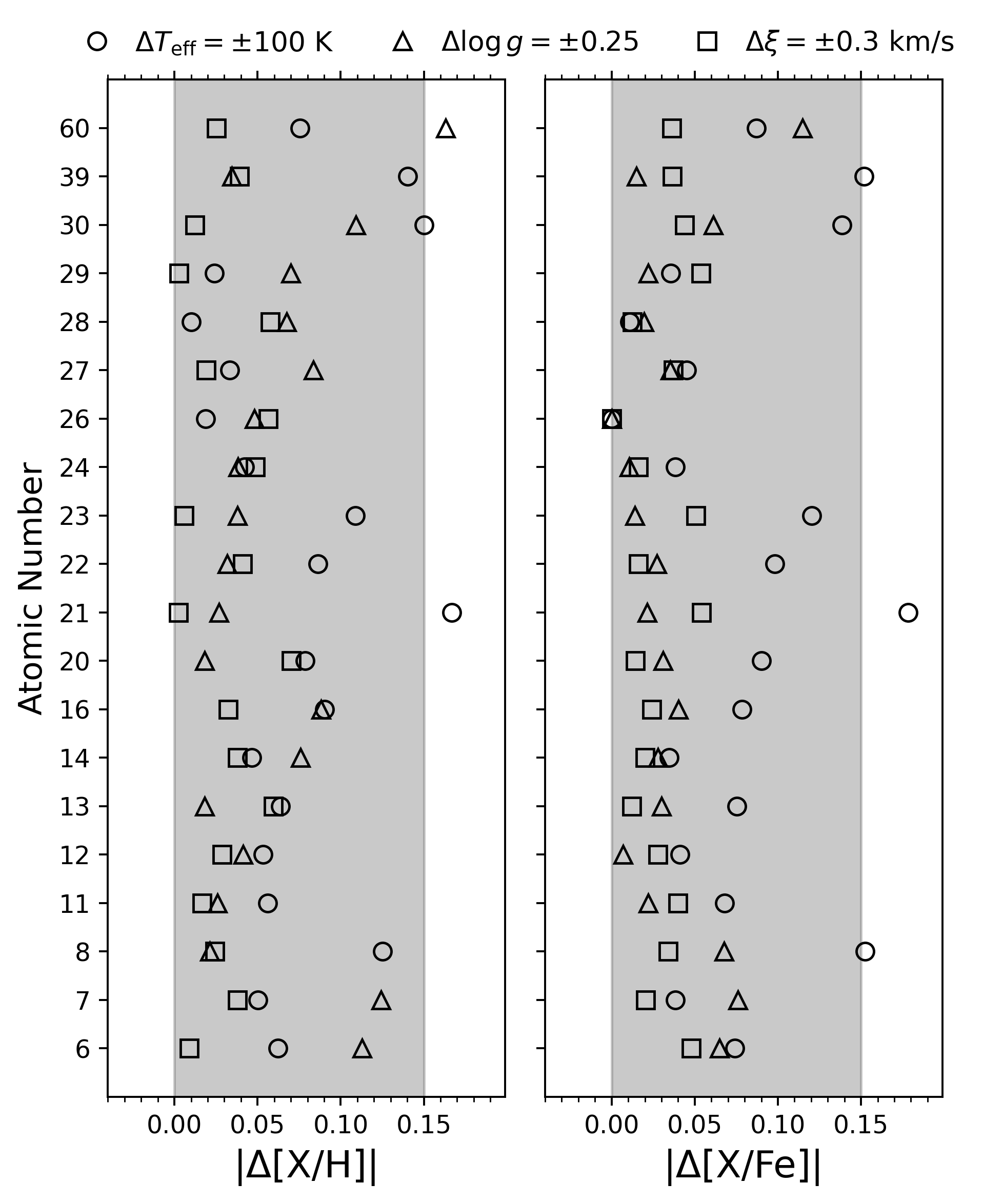}
   \caption{ Absolute values of the typical differences in the derived [X/H] and [X/Fe] with varying T$_{\rm eff}$ by $\pm$100~K (circles), log~g by $\pm$0.2~dex  (triangles) and $\xi$ by $\pm$ 0.3~km~s$^{-1}$ (squares) for each measured chemical element, identified by its atomic number. The grey shaded area indicates the region where differences are within 0.15 dex.}
    \label{fig_errors}
\end{figure}

\subsection{Atmospheric parameters}
For an accurate application of spectral synthesis, the values of T$_{\rm eff}$ and log(g) for all the observed stars are needed.
As already successfully done in the previous works \citep{deimer24,fanelli+24,ferraro+25a}, we obtained first-guess estimates of these parameters from the projection of each target onto the closest between the two selected isochrones in the differential reddening-corrected CMD, under the assumption of a distance modulus $(m - M)_0 = 14.65$ and an average color excess of $E(B - V) = 4.52$ \citep{pallanca_21, ferraro_21}. 
The two isochrones have been selected from the PARSEC database \citep{bressan_12,marigo+17} according to the properties of the two main sub-populations of Liller~1(see \citealt{ferraro_21} and \citealt{dalessandro_22}): one has an age of 12 Gyr and an iron abundance [Fe/H]$= -0.3$, the other is 2 Gyr old with [Fe/H]$ = +0.3$. 
A spectroscopic fine-tuning of the atmospheric parameters has then been obtained through the simultaneous fit of the OH and CO molecular lines and band heads.

 \begin{figure}[]
\centering
    \includegraphics[width=\columnwidth]{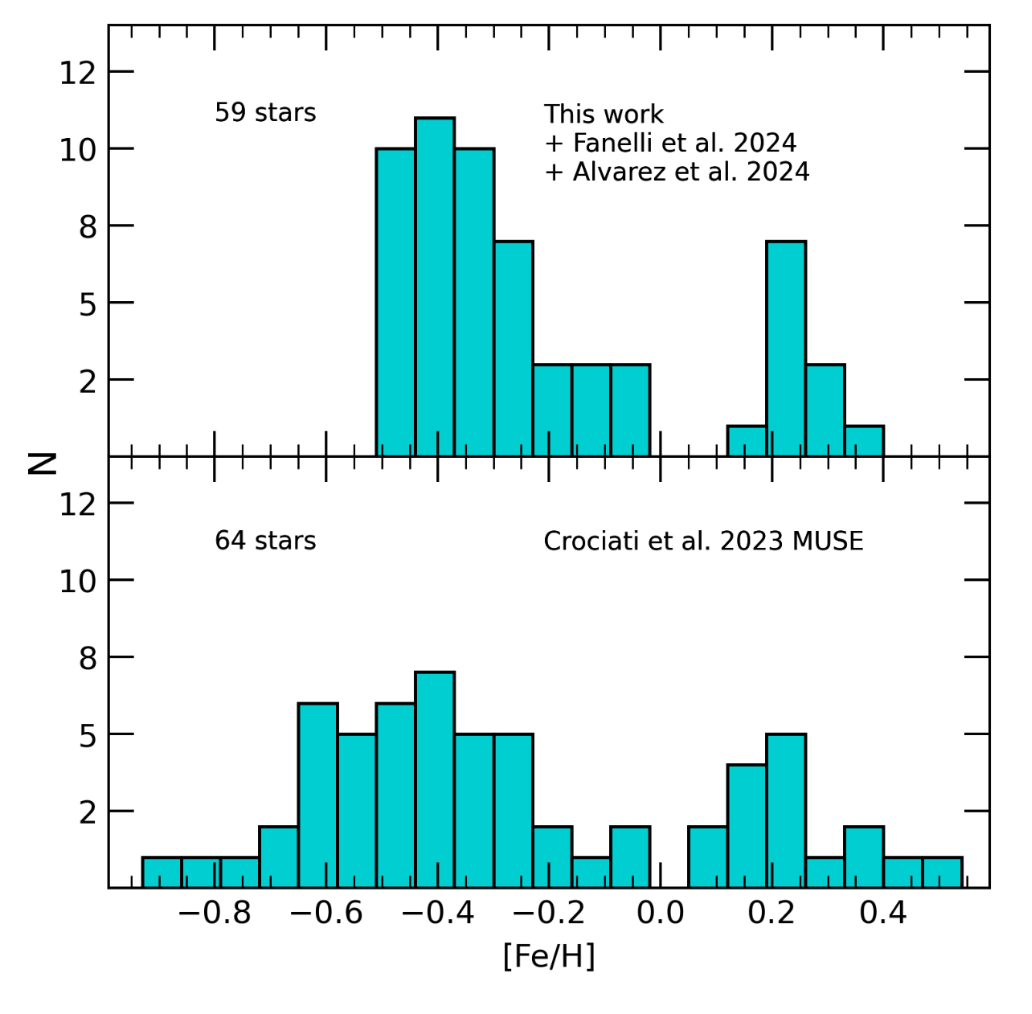}
    \caption{ [Fe/H] distribution obtained from this work and previous measurements from \citet{deimer24} and \citet{fanelli+24} (top panel), and that of \citet[][bottom panel]{crociati_23}. }
    \label{isto}
\end{figure}

\subsection{Radial velocities}
To determine the heliocentric radial velocities (RVs) of the 30 target stars, we cross-correlated the observed spectra with the synthetic ones. The resulting values (see Table \ref{tab1} and bottom-right panel of Fig. \ref{fig1}) range from $51.8$ to 86.2 km s$^{-1}$, with a typical uncertainty of smaller than 1 km~s$^{-1}$, a mean value of (68.7$\pm$ 1.5) km~s$^{-1}$ and a 1$\sigma$ dispersion of (8.1$\pm$1) km~s$^{-1}$. The mean RV and dispersion of Liller 1 obtained from previous studies \citep{crociati_23, deimer24, fanelli+24} are 66.7 km s$^{-1}$ and 8.7 km s$^{-1}$, respectively (see \citealp{ferraro+25a}). Hence, we conclude that all the 30 measured stars have RVs within 2$\sigma$ from the systemic velocity, supporting their membership to Liller 1, as already indicated by proper motions. 

\begin{figure*}[]
        \centering
        \includegraphics[width=\textwidth]{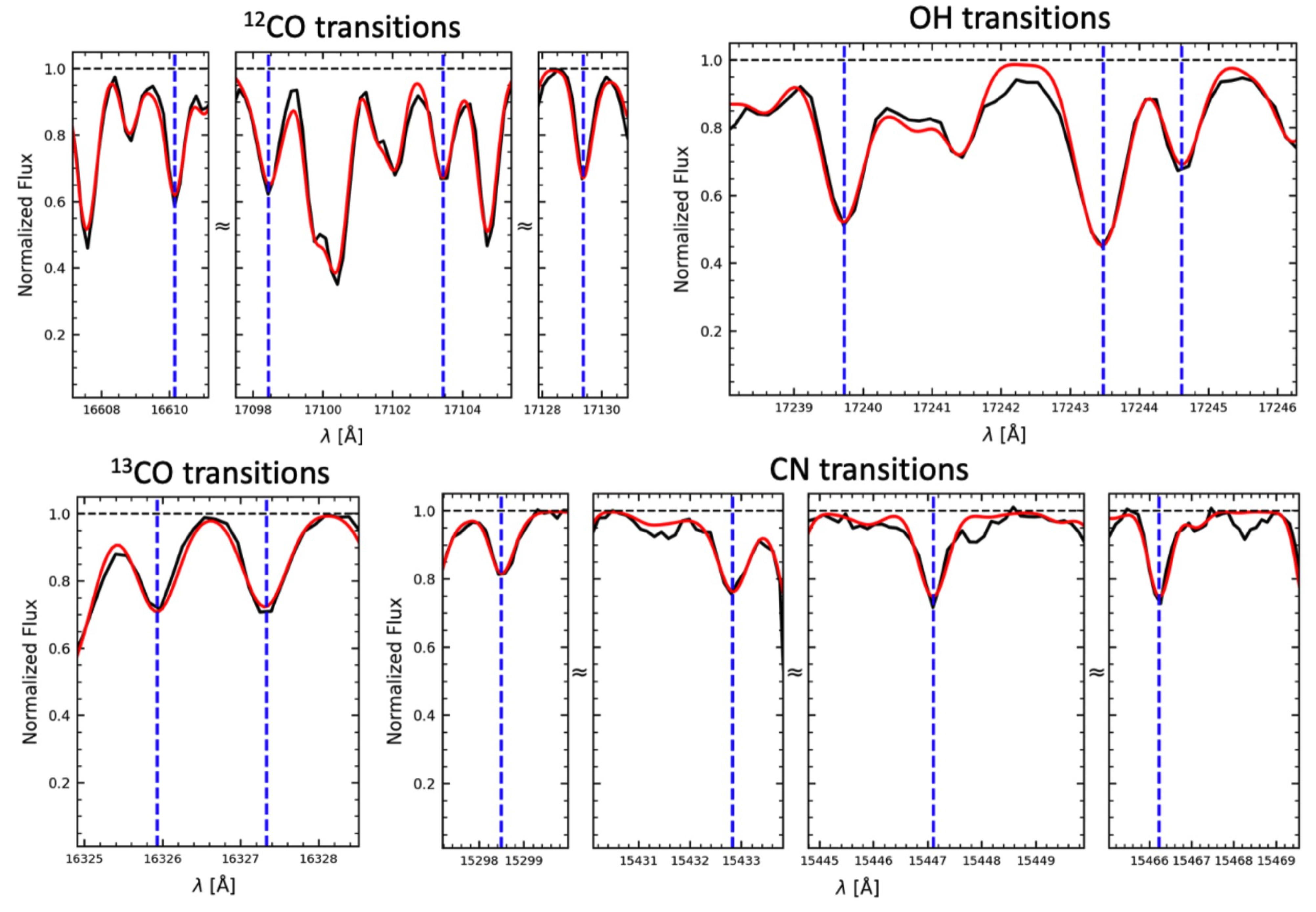}
         \caption{ Examples of  $^{12}$CO, $^{13}$CO, OH, and CN molecular transitions used in the present chemical analysis (marked by the vertical dashed lines). In each panel, the observed spectrum (black line, ID 387099) and the corresponding synthetic fit (red line) are shown.
         }
\label{fig_spettri}
\end{figure*}

\subsection{Chemical abundances}
\label{chemabu}
The detailed chemical abundances of Fe, C, N, O, Na, Mg, Al, Si, S, Ca, Sc, Ti, V, Cr, Co, Ni, Cu, Zn, Y and Nd for the 30 observed stars in Liller~1 were determined through spectral synthesis of atomic and molecular lines. 
While the full list of lines used for such a chemical analysis is provided in machine-readable format,  Table~\ref{tab2} reports the first rows, 
for the sake of illustration.
We first derived the abundances of iron and $\alpha$-elements from atomic lines. 
Then, to minimize possible degeneracies in the measurements of individual CNO abundances from molecular lines, an iterative spectral synthesis approach was applied to each star, by fixing the metallicity according to the inferred iron and $\alpha$-element abundances,  and moving C first in order to fit the $^{12}\mathrm{C}$O molecular lines, and then O and N to fit the OH and CN molecular lines, respectively,  until a reasonable convergence was achieved. Once the C abundance has been fixed, the $^{12}\mathrm{C}$/$^{13}\mathrm{C}$ abundance ratio was also obtained by means of spectral synthesis of the $^{13}\mathrm{C}$O molecular lines. Finally, we derived the abundances of all the other elements.

\begin{figure}[t]
        \centering
        \includegraphics[width=\columnwidth]{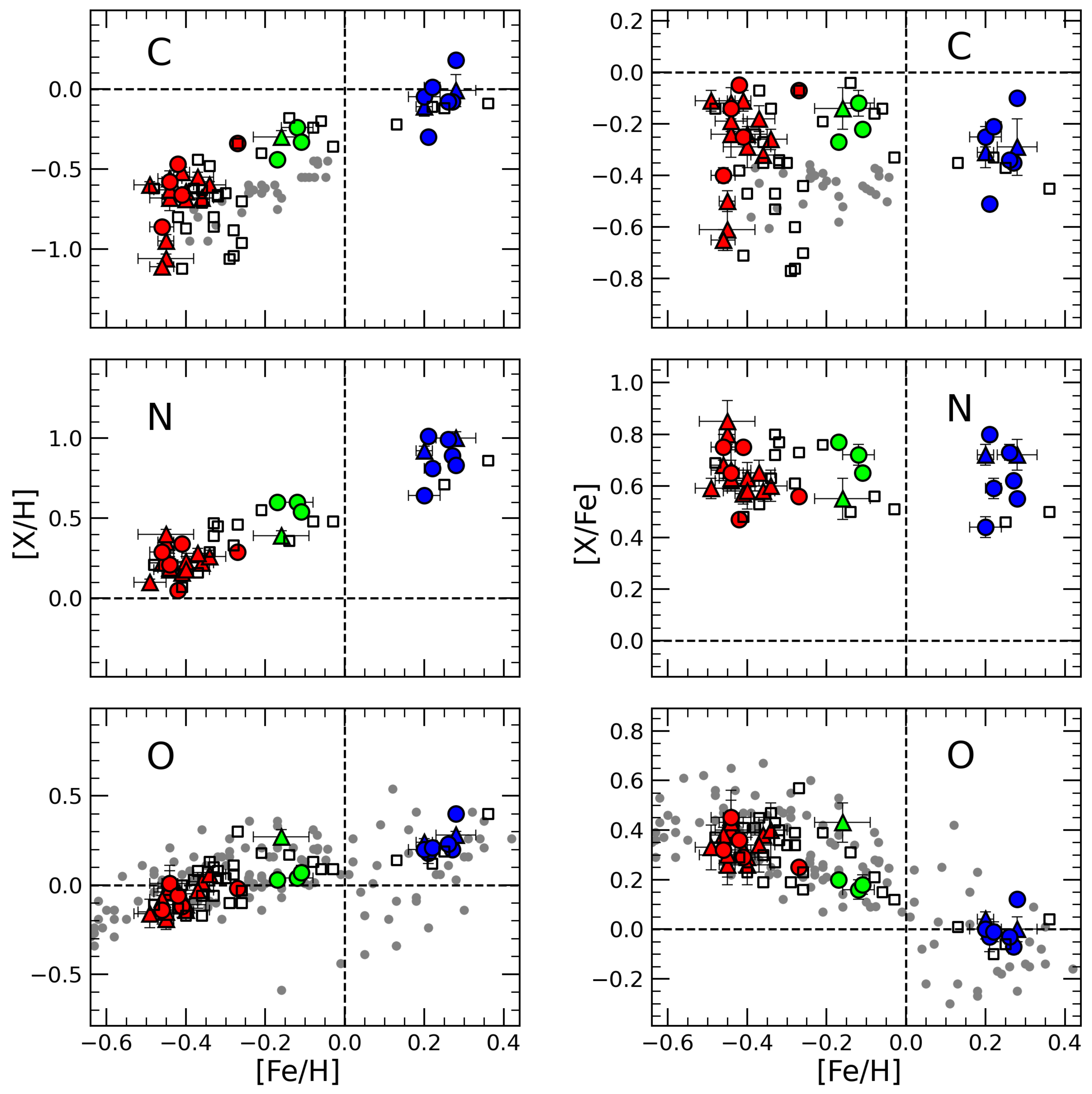}
         \caption{[X/H] (left panels) and [X/Fe] (right panels) of CNO elements as a function of [Fe/H], for the 30 stars of Liller~1 observed with CRIRES+. Red symbols mark metal-poor stars (with [Fe/H]$<-0.2$), green symbols correspond to the metal-intermediate stars (with $-0.2<$[Fe/H]$<0$), and blue symbols represent the metal-rich stars (with [Fe/H]>0). Triangles refer to stars measured in program 109.230K (hereafter, P109), while the large circles are the stars measured in program 110.24A4 (hereafter, P110; see Sect.~\ref{obs} for more details). The empty squares are previous measurements from \citet{deimer24} and \citet{fanelli+24}. The light gray circles are measurements for reference bulge field stars from \citet{rich12} and \citet{johnson_14}. The dashed vertical and horizontal lines mark the solar values.}
\label{fig_CNO}
\end{figure}

\begin{figure}[t]
        \centering
        \includegraphics[width=\columnwidth]{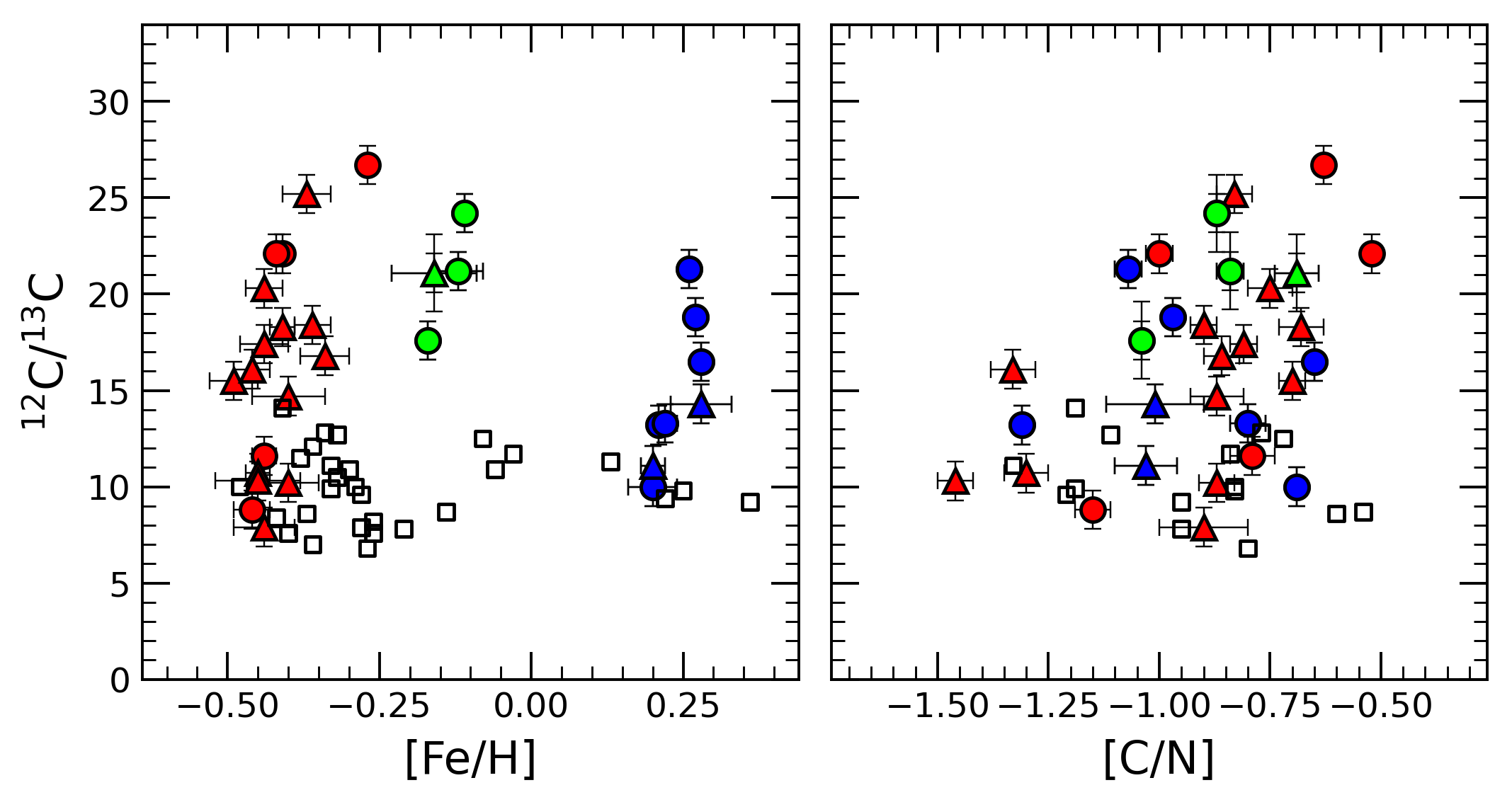}
    \caption{Isotopic ratio $^{12}\mathrm{C}$/$^{13}\mathrm{C}$ as a function of [Fe/H] (left panel) and [C/N] (right panel) for the 30 stars of Liller~1 observed with CRIRES+. The meaning of different colors and symbols is as in Fig. \ref{fig_CNO}.}
    \label{fig_C12C13}
\end{figure}

\begin{figure*}[]
\centering
        \includegraphics[width=\textwidth]{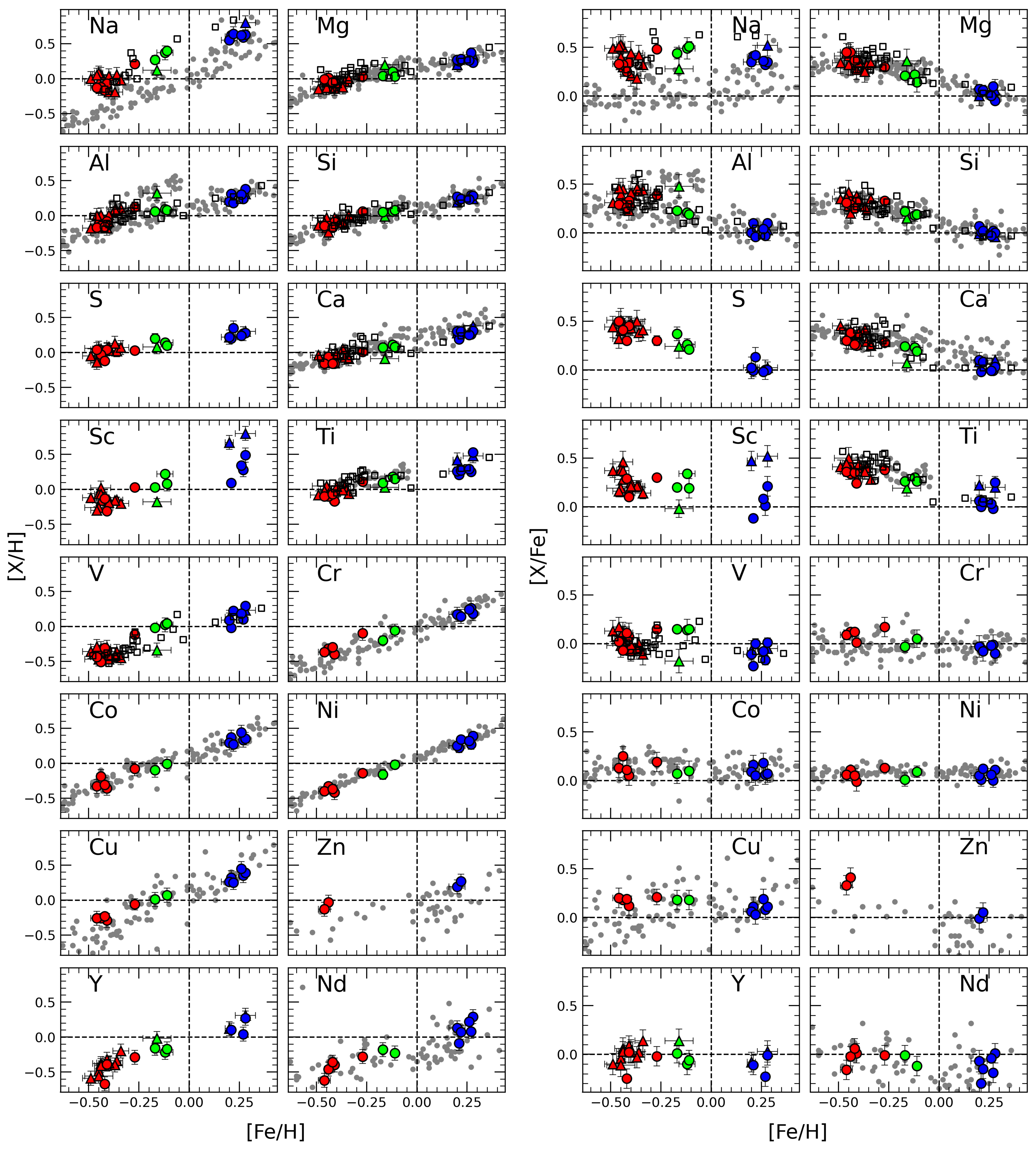}
    \caption{[X/H] (left panels) and [X/Fe](right panels) of alpha, Al, Na, iron-peak and neutron-capture elements as a function of [Fe/H] for the 30 stars of Liller~1 observed with CRIRES+. The meaning of different colors and symbols is as in Fig. \ref{fig_CNO}, with the only exception of the bulge field stars (light gray circles) that in the Zn panel are from \citet{barbuy_15}, and in the Nd panel are from from \citet{johnson12} and \citet{swaelmen16}.}
\label{fig_abun}
\end{figure*}

The error budget of the derived [X/H] abundances and [X/Fe] abundance ratios envisages the contribution of random and systematic uncertainties.
Random uncertainties in the derived abundances primarily arise from errors in continuum placement (typically  1–2\%) and from photon noise. For elements with multiple measurable lines, such stochastic errors have been evaluated as the standard deviation divided by the square root of the number of used lines. For elements with a single measurable line, a conservative uncertainty of 0.10 dex was assumed. The random uncertainties are reported in Table 2 and plotted as error bars in Figs.~\ref{fig_CNO}-\ref{fig_abun}.
Systematic uncertainties mostly arise from stellar parameters, and they typically amount to $\pm$50-100 K in T$_{\rm eff}$, $\pm$0.2 dex in log(g) and $\pm$0.3 km s$^{-1}$ in microturbulence velocity. We performed specific tests to evaluate the impact of these uncertainties on the derived [X/H] abundances and [X/Fe] abundance ratios in the range of stellar parameters and metallicities covered by the observed Liller~1 stars. The inferred abundance differences do not show any significant trend with the stellar parameters themselves and/or the metallicities covered by the measured stars, nor significant asymmetries between positive and negative variations of the stellar parameters. Hence, we computed an average value for each parameter variation. 
These abundance variations are summarized in Fig.~\ref{fig_errors} for each measured chemical element. As can be appreciated, the derived $\Delta$[X/H] and $\Delta$[X/Fe] are normally well within 0.15 dex, often within 0.1 dex. It is also worth noticing that in real giant stars, variations of these stellar parameters are not independent, and they can partially compensate each other in terms of impact on the derived abundances, thus reducing the resulting $\Delta$[X/H] and $\Delta$[X/Fe]. We finally highlight that systematic uncertainties in the stellar parameters have mostly the effect of rigidly shifting the chemical distributions. 
However, since these shifts turn out to be relatively small (if any) in the space of parameters covered by the measured stars in Liller 1, they do not significantly affect the overall appearance of the chemical distributions.

\subsubsection{Iron abundance distribution}

The iron abundance distribution of Liller~1 shows two main components: a sub-solar one, counting 18 stars with an average [Fe/H]=$-0.42\pm 0.01$ dex and a 1$\sigma$ dispersion of 0.05$\pm 0.01$ dex, and a super-solar population, counting 8 stars, with an average [Fe/H]= +0.23$\pm 0.01$ dex and a 1$\sigma$ dispersion of 0.04$\pm 0.01$ dex. In addition, four stars have been found at intermediate metallicities, in the range $-0.2<$[Fe/H]$<0.0$, with an average value of $-0.14\pm 0.02$ dex and a 1$\sigma$ dispersion of 0.03$\pm 0.01$ dex.

This metallicity distribution is consistent with previous results reported in \citet{crociati_23}, \citet{deimer24},  \citet{fanelli+24} and \citet{ferraro+25a}, with the expectations derived from the reconstructed star formation history of the system \citep{dalessandro_22}, and with what observed in Terzan 5 \citep{origlia+25}. Fig.~\ref{isto} summarizes the results obtained so far by our group, showing the metallicity distribution derived from high- and medium-resolution spectroscopy (top panel), and that found from the analysis of MUSE spectra (bottom panel). For the stars in common with previous samples, the CRIRES+ results are adopted. 

\subsubsection{CNO}
The chemical abundances of C, N, and O were derived from dozens of CO, CN, and OH molecular transitions, respectively. In particular, the $^{12}\mathrm{C}$O lines were used to measure the abundance of carbon and, together with the $^{13}\mathrm{C}$O lines, to determine  the carbon $^{12}\mathrm{C}$/$^{13}\mathrm{C}$ isotopic ratio, reported in the last column of Tables \ref{tab3} and \ref{tab4}.
For the sake of illustration, a few examples of the used molecular transitions
are shown in Fig. \ref{fig_spettri}.
A few lines of atomic carbon have been also measured as a sanity check of both the inferred C abundances and stellar temperatures \citep{fanelli21}, finding fully consistent results.

Fig.~\ref{fig_CNO} displays the [C/H], [N/H] and [O/H] abundances and the [C/Fe], [N/Fe] and [O/Fe] abundance ratios as a function of [Fe/H]. 
[C/Fe] is depleted relative to the solar-scaled value in all the stars, and the scatter is noticeably larger at [Fe/H]$<-0.2$ (with an overall 1$\sigma$ dispersion of about 0.2 dex) than in the other cases.
[N/Fe] is enhanced relative to the solar-scaled value in all the stars, with an overall 1$\sigma$ dispersion of about 0.1 dex across the three sub-populations. 
[O/Fe] is enhanced by 0.2-0.4 dex in the metal-poor and intermediate-metallicity populations, while it reaches solar-scaled values in the metal-rich one.

We also measured $^{12}\mathrm{C}$/$^{13}\mathrm{C}$ isotopic ratios in the 8-27 range, with the highest values observed  in the warmest stars, and with a typical uncertainty of $\pm$1.
No significant trend of $^{12}\mathrm{C}$/$^{13}\mathrm{C}$ with metallicity or with the [C/N] abundance ratio has been found, as can be seen in Fig.~\ref{fig_C12C13}.
The inferred CNO abundances and the low ($<15$) $^{12}\mathrm{C}$/$^{13}\mathrm{C}$ isotopic ratios measured in the most luminous giants are fully consistent with previous measurements by \citet{deimer24} and \citet{fanelli+24}.

\subsubsection{Iron-peak elements}
Regarding iron-peak elements, Sc was measured in the $K$ band, V in both the $H$ and $K$ bands, while Cr, Co, Ni, and Cu were derived from absorption lines in the $H$ band. 
Zn was measured from the only available line in the $J$ band and just for the four, most luminous stars that were also observed in this band. 
As shown in Fig. \ref{fig_abun}, the [X/Fe] values for Cr, Co, Ni, and Cu are about solar-scaled or slightly enhanced, and they exhibit a flat trend with varying [Fe/H], following the behavior observed in bulge field stars.
The [V/Fe] distribution shows some scatter around the solar-scaled value, except for two stars that have [V/Fe] values significantly depleted (by more than a factor of two) compared to the solar-scaled value.
[Sc/Fe] shows some enhancement with respect to the solar-scaled value and some spread. Two metal-rich stars, in particular, have especially strong Sc lines, whose modeling is not trivial, as also discussed by \citet{Thorsbro18}, \citet{Thorsbro20_1}, and \citet{Thorsbro20_2}. 
[Zn/Fe] abundance ratios were measured in only two metal-poor and two metal-rich stars, finding enhanced (by about 0.4 dex) and approximately solar-scaled values, respectively.

\subsubsection{$\alpha$-elements, Ti, Na and Al}
Fig.~\ref{fig_abun} displays the [X/H] abundances and the [X/Fe] abundance ratios as a function of [Fe/H] also for Mg, Si, S, Ca, and Ti. All these elements exhibit a broadly similar behavior.
The metal-poor component shows enhanced (by a factor of 2-3) abundance ratios, while the metal-rich component displays about solar-scaled values. The metal-intermediate subpopulation exhibits a lower enhancement than that of the metal-poor one. Within the errors,  these results are consistent with previous spectroscopic studies of Liller~1 \citep{deimer24,fanelli+24,ferraro+25a}, and they also closely match the chemical pattern shown by the bulge field stars (small grey circles).
Fig. \ref{fig_abun} also shows the results for Al and Na.
The behavior of [Na/Fe] as a function of [Fe/H] is similar across the three subpopulations, showing values that are enhanced by a factor of 2-3 with respect to the solar-scaled one. In contrast, Al behaves like the $\alpha$-elements, showing enhanced [Al/Fe] in the metal-poor and intermediate-metallicity stars, and approximately solar-scaled values in the metal-rich ones.

\subsubsection{Neutron-capture elements}
Compared to previous spectroscopic studies, this work extends the chemical analysis to an additional class of elements, namely the neutron-capture species. In particular, we were able to measure the abundances of Y and Nd, both of which are mainly produced through the s-process, with s/r ratios of approximately 70/30 and 60/40, respectively \citep{Bist14,Prant20}. We measured Y from one line in the $K$ band, and Nd from a couple of ionized lines in the $H$ band. As shown in Fig.~\ref{fig_abun}, these elements display a very similar behavior with a mild (if any) trend of [X/Fe] vs [Fe/H], with the sub-solar stars being about solar-scaled, and the super-solar ones being slightly depleted with respect to the solar-scaled value. 
In the case of Nd, some measurements of bulge stars from high-resolution optical spectroscopy \citep[see e.g.][]{johnson12,swaelmen16} and from APOGEE \citep{salessilva24} are also available, showing a [Nd/Fe] trend with metallicity fully consistent with the one found in Liller~1.

\section{Comparison with other studies}
\label{comp}
The ten brightest targets studied in this work are in common with the samples of \citet{deimer24} and \citet{fanelli+24}. We found an overall agreement among the measured abundances for the eleven chemical elements in common, with average differences and dispersions that do not exceed 0.1 dex, thus also providing a high-resolution validation of 
those two previous studies that were based on lower-resolution spectra.

Very recently, an APOGEE-based analysis of Liller~1 has been presented by \citet{liptrott26}, claiming significant chemical differences between Liller 1 and the bulge, and suggesting  a possible extra-galactic origin for this stellar system.
One star, 2MASS J17332478-3326071 (ID 387099), is in common between their sample and the CRIRES+ one. For this star, the adopted stellar parameters are fully consistent between the two analyses within the quoted uncertainties, and the chemical abundances are in agreement within 0.1-0.2 dex, with the only exception of C and especially N, for which we find an offset of $\sim 0.2$ and 0.4, respectively. 
Although systematic differences from studies that use different datasets and modeling are somewhat physiological, it is worth noticing 
that in our study, thanks also to the high spectral resolution of CRIRES+, we derived C abundances from dozens of roto-vibrational CO transitions (see Table~\ref{tab2}) carefully selected to be free from major contamination by other species. \citet{liptrott26} seems to have used different CO transitions (see, e.g., their Fig. 3), among those that we have discarded since significantly affected by blending with other species, thus possibly leading to systematically higher values of C.
As a consequence of the molecular equilibria, such a C over-abundance propagates into a corresponding N under-abundance, thus likely explaining the observed C–N offset between the two works.

\begin{figure}[t]
        \centering
        \includegraphics[width=\columnwidth]{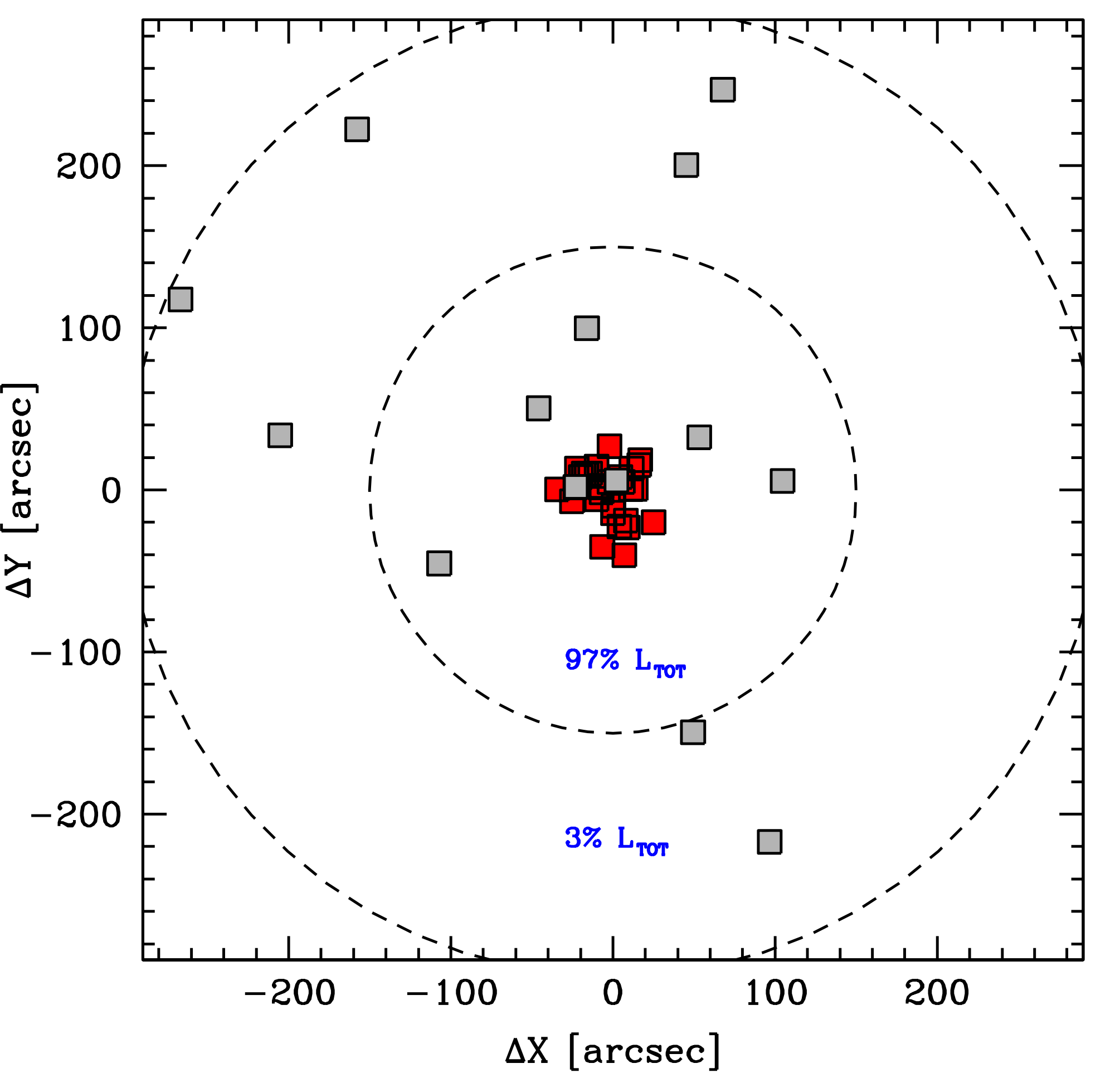}
    \caption{Spatial distribution of the stars discussed in this work (red squares) and those discussed in \citet{liptrott26} (gray squares). The inner dashed circle is at 150" from the center and includes 97\% of the Liller 1 total light. The outer dashed circle corresponds to the tidal radius ($r_t=298"$) quoted in \citet{saracino_15}.}
    \label{fig_lip}
\end{figure}

Overall, the work of \citet{liptrott26} is based on the analysis of APOGEE spectra for 14 very bright ($K<10$) giant stars, half of which are located in the outskirts of Liller~1  (see Figure \ref{fig_lip}), between  $150''$ (corresponding to five times the half-mass radius and 16 times the core radius) and the cluster tidal radius ($r_t\sim 300''$; \citealp{saracino_15}), i.e., in a region where the surface brightness (hence the  sampled luminosity) is overwhelmed by the Galactic field contribution (see Fig. 8 in \citealp{saracino_15}).

By assuming the King model that best fits the surface brightness profile of Liller~1 (see \citealp{saracino_15}), the cluster luminosity sampled between $150''$ and $300''$ is only 3\% of the total, corresponding to $L_s=2\times 10^4 L_\odot$. On the other hand,
using BASTI evolutionary tracks \citep{pietrinferni+13}, we find that the evolutionary time required by a Liller~1 star to evolve along the brightest portion ($K<10$) of the RGB is just 3 Myr.
With these data, we estimated the number of stars ($N$) expected to be observable in the bright portion of the RGB and in the radial range $150''<r<300''$ from the center of Liller~1, by using the well-known relation $N = 2 \times 10^{-11} \times L_s \times t$ \citep{renzini86}, where $L_s$ is the sampled luminosity  and  $t$ is the evolutionary time. 

We find that only one star is expected, while \citet{liptrott26} claimed for the presence of seven member stars.  This indicates that a significant fraction of their targets are likely not members, in spite of their radial velocities and proper motions.\footnote{\citet{liptrott26} assume as likely cluster members all the stars with proper motion within five times the mean proper motion dispersion of Liller 1, and with radial velocity differing by less than two times the central velocity dispersion of the system. These seem to be quite generous criteria, especially for stars orbiting in the cluster periphery, where the velocity dispersion significantly drops with respect to the central value.}
In this respect, it is worth highlighting that Liller~1 (like Terzan~5) is a compact and heavily field-contaminated system, with radial velocity and proper motion distributions that largely overlap with those of the surrounding field (see, e.g., Fig. 2 in \citealp{ferraro_21}).

Independently of how many stars of the \citet{liptrott26} sample are actually Liller~1 members, they are too few to trace the complex, multi-age and multi-metallicity populations of Liller~1, and to draw  conclusions on the origin of this system \citep[see also][]{origlia+25}. 
Indeed, while APOGEE is a very powerful instrument to investigate the chemistry of stellar systems at relatively large spatial scales and low stellar density, it is not very effective in exploring compact, high-density, and heavily contaminated systems like Liller~1 and Terzan~5, 
being limited to the outer regions, where the field contamination is large,  and the probability of selecting truly member stars drops significantly.
For a reliable characterization of these systems and to properly constrain their formation scenarios, instead, it is mandatory to select a sufficiently large number of targets in the innermost regions.
Our CRIRES+ sample counts 30 stars all distributed within 40'' (i.e, within about one half-mass radius) and with kinematics consistent with the system bulk motion, thus maximizing the probability of their membership. These stars have chemical patterns that are fully consistent with those measured in the bulge. 
If the chemical anomalies claimed by \citet{liptrott26} with respect to the bulge are confirmed, they could possibly trace a different sub-structure not connectable to Liller~1 but with a potential extra-Galactic origin.

\section{Discussion and conclusions}
\label{disc}

The analysis of the high resolution CRIRES+ spectra for 30 member stars in Liller~1 provided us with accurate abundances of 19 different chemical elements.
In our study, carbon and nitrogen abundances and abundance ratios, as well as the $^{12}\mathrm{C}$/$^{13}\mathrm{C}$ isotopic ratios show the typical signatures of mixing and extra-mixing processes occurring in the stellar interiors during the evolution along the RGB \citep{cha95,den96,cav98,boo99}. 
Concerning iron-peak elements, V and Cr closely trace iron, while Sc, Co, Ni and Cu show some enhancement with respect to the solar-scaled values. The [X/Fe] abundance ratios of these iron-peak elements show no significant trend with metallicity 
and they are consistent with the presence of self-enriched sub-populations with different ages and metallicity. Measurements of Zn abundance are available for four stars only, and follow the trend observed of the bulge field stars, with some enhancement at sub-solar metallicity and about solar-scaled values at super-solar one.
Globally, the iron-peak abundance distributions of Liller~1 closely match those of the bulge field. The two measured neutron-capture elements (Y and Nd) show a mild trend with metallicity around the solar-scaled values, consistent with that of the bulge field.

\begin{figure}[t]
    \centering
    \includegraphics[width=\columnwidth]{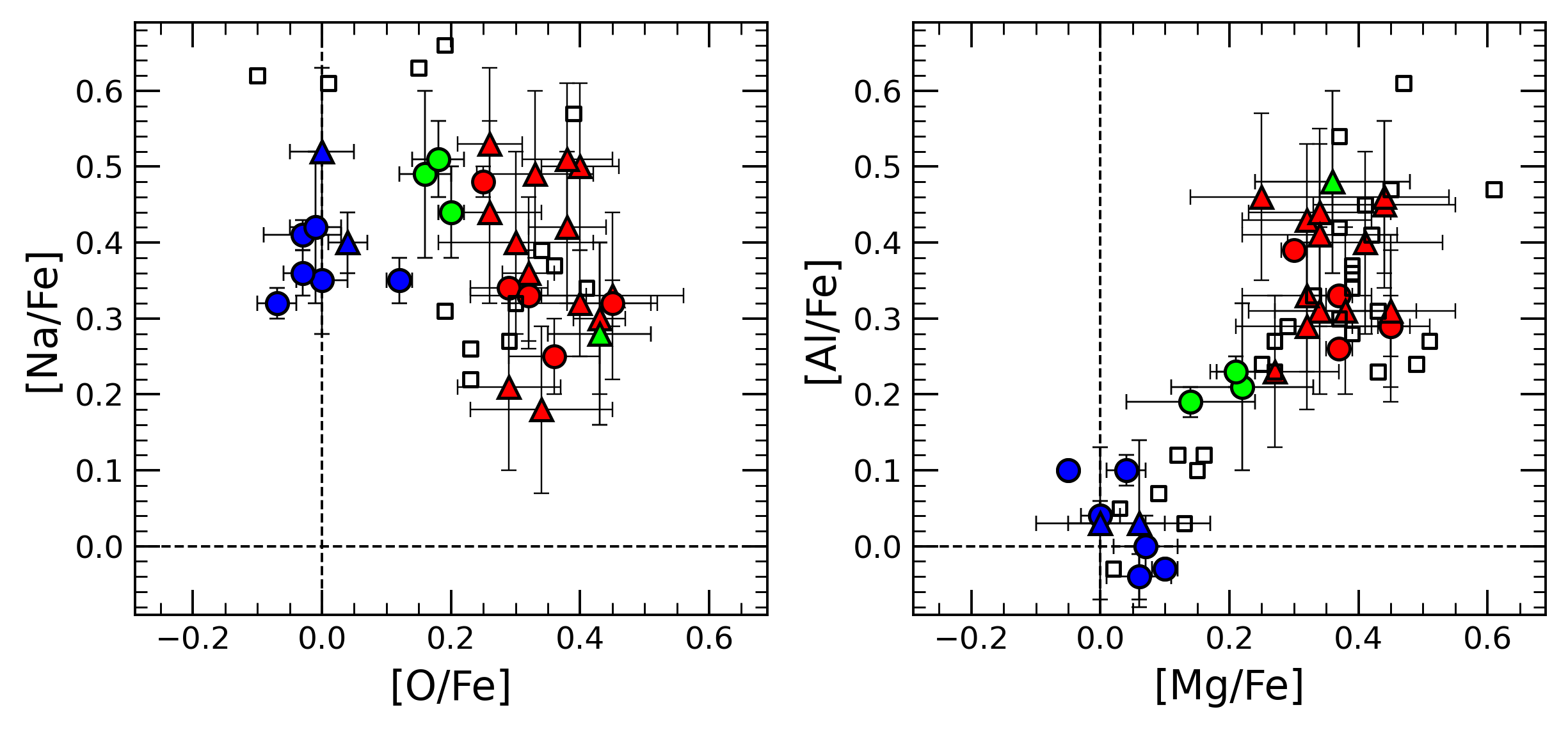}
   \caption{ [Na/Fe] as a function of [O/Fe] (left panel) and  [Al/Fe] as a function of [Mg/Fe](right panel) for the 30 stars of Liller~1 observed with CRIRES+. The meaning of different colors and symbols is as in Fig. \ref{fig_CNO}}
    \label{fig_AlNa}
\end{figure}

\begin{figure}[t]
    \centering
    \includegraphics[width=\columnwidth]{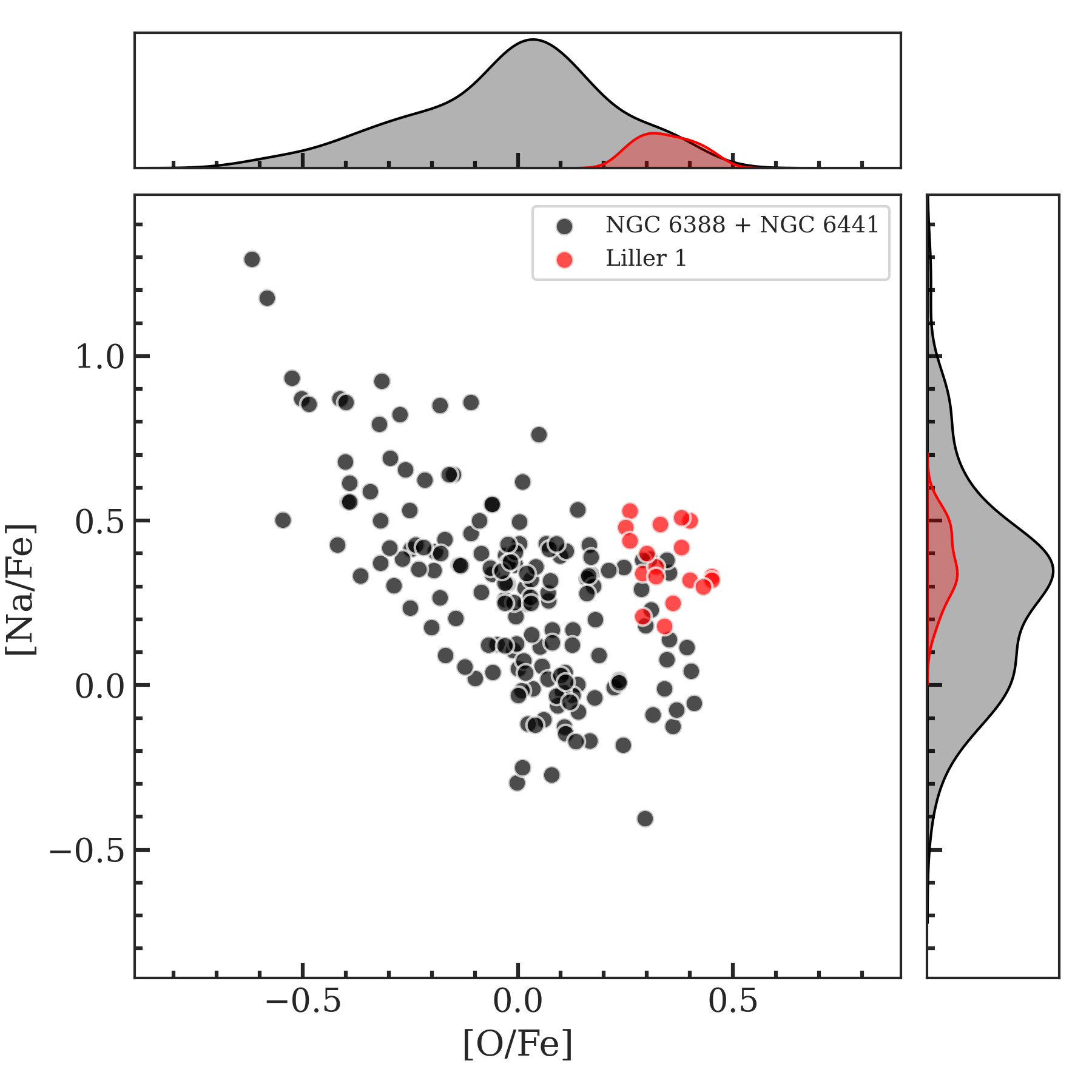}
   \caption{Distribution of the dominant, old and metal-poor ([Fe/H]$=-0.4$) component of Liller 1  (red circles) in the [Na/Fe]-[O/Fe] diagram, compared to that of a control sample made of two genuine GCs with similar metallicity (namely, NGC 6441 and NGC 6388; black circles, from \citealp{gratton+07, carretta+23}). At odds with the control clusters, Liller 1 stars do not exhibit the typical O-Na anti-correlation characteristics of genuine GCs. The kernel density estimation curves highlight this disparity, showing a much narrower oxygen and sodium spread for Liller 1 (red shaded), compared to the broad distribution of the comparison dataset (grey shaded).}
    \label{fig_NaO}
\end{figure}

\begin{figure*}[t]
\centering
    \includegraphics[width=\textwidth]{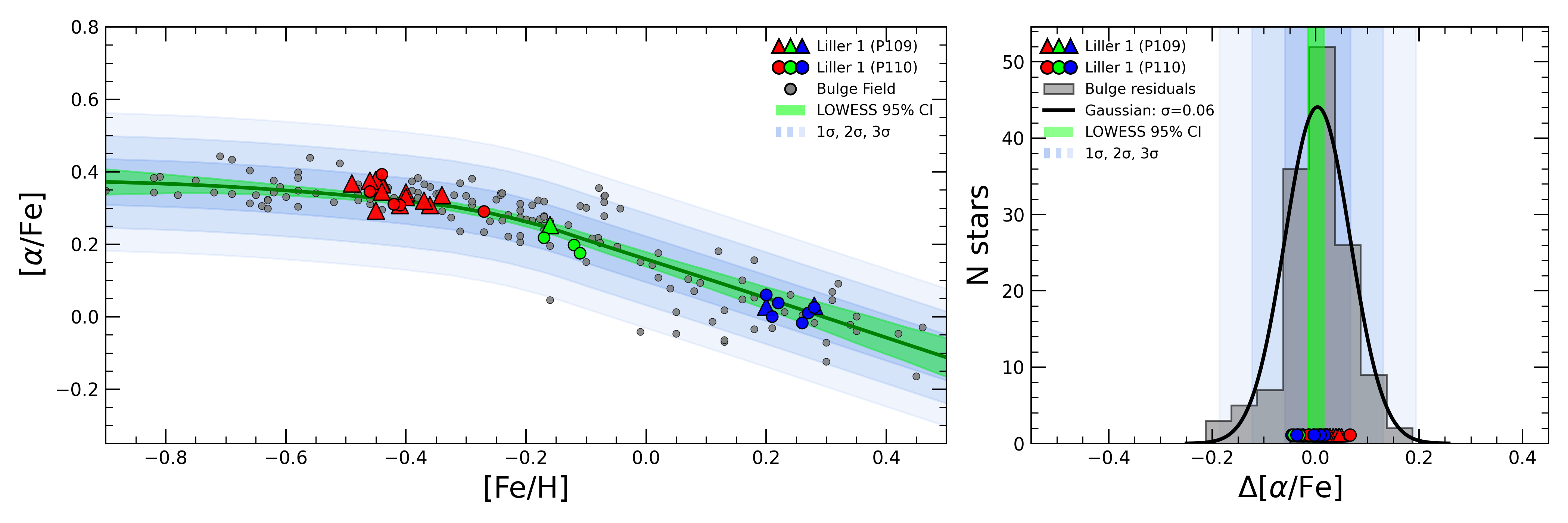}
    \caption{Behavior of the mean [$\alpha$/Fe]  abundance ratio (from O, Mg, Si and Ca) as a function of [Fe/H] for the 30 stars observed in Liller~1 in this work. Left panel: the large colored symbols mark the observed targets, with the same meaning as in Fig. \ref{fig_CNO}. Bulge field stars (from \citealt{rich12} and \citealt{johnson_14}) are also shown as gray circles for comparison. The 95\% confidence confidence region around the LOWESS median trend of the bulge distribution is marked by the green band, while the 1, 2, and 3$\sigma$ confidence regions are indicated by progressively lighter gray  shaded stripes. Right panel: distributions of the residuals from the LOWESS median for both the bulge field stars (gray histogram and its Gaussian fit) and for the Liller~1 stars (colored symbols and vertical bands with the same meaning as in the left panel).}
    \label{statistic_alpha}
\end{figure*}

All the investigated chemical elements, including iron, span a wide range (from 0.5 to about 1 dex) of abundance values (see the left panels in Fig. \ref{fig_abun}), indicating a complex chemical evolution and a formation pathway different from that of genuine GCs. However, a few scenarios proposed in the literature for the formation of Liller 1 and Terzan 5 include the merger between two GCs \citep{pfeffer21}, or the accretion of a giant molecular cloud by a genuine GC \citep{mckenzie18,bastian_22}. Hence, we also investigated the possibility that some of the Liller 1 sub-populations, in particular the dominant, old and sub-solar ([Fe/H]$\sim-0.4$) component, show the light-element anti-correlations characteristics of genuine GCs (see, e.g., \citealt{carretta09}). As apparent from Fig.~\ref{fig_AlNa}, no evidence of Na-O or Al-Mg anti-correlations is found in any of the Liller 1 sub-populations. In addition, the  Spearman rank test performed on the [Na/Fe]-[O/Fe] distribution of the oldest component yields a correlation coefficient $\rho= -0.3$ with a $p$-value of 0.22. A bootstraping procedure based on 10,000 extractions demonstrated that this value of the correlation coefficient is consistent with zero within the $95\%$ confidence interval. 
This clearly indicates that no statistically significant correlation is detected. We also compared (see Fig. \ref{fig_NaO}) the [Na/Fe]-[O/Fe] distribution of the Liller1 sub-solar component with those observed in NGC 6388 and NGC 6441 \citep{carretta+23,gratton+07}, two GCs with similar [Fe/H] abundance (because the extension of this anti-correlation is known to be sensitive to metallicity; see, e.g., \citealp{carretta09}). The 2D Kolmogorov-Smirnov test yields a $p$-value $<0.01$, strongly rejecting the null hypothesis that the dominant population of Liller 1 and the comparison sample are drawn from the same parent distribution, and therefore suggesting that this system does not derive from a GC.

To get deeper insights into the complex formation and chemical enrichment history of Liller~1 and following the analyses performed in \citet{ferraro+25a} and in \citet{origlia+25},  we computed the mean abundance of  
O, Mg, Si and Ca ($\alpha$-elements) for the Liller 1 stars measured in this work, and for more than 170 bulge field stars measured by \citet{rich12} and \citet{johnson_14} through high resolution optical and NIR spectroscopy.
The left panel of Fig. \ref{statistic_alpha} shows the resulting [$\alpha$/Fe] versus [Fe/H] distribution. We fitted the bulge distribution with a locally weighted scatterplot smoothing\footnote{\url{https://www.statsmodels.org/devel/generated/statsmodels.nonparametric. smoothers lowess.lowess.html}} (LOWESS) regression, which provides a non-parametric estimate of the median trend (see the green curve in the left panel of Fig. \ref{statistic_alpha}). The right panel presents the distributions of the vertical deviations from the LOWESS curve for both the bulge  (gray histogram) and the Liller~1 stars (large colored symbols). As apparent, Liller~1 stars fall comfortably  within the intrinsic bulge dispersion, with 100\% of them being inside 1$\sigma$. These statistics demonstrate the chemical consistency between Liller~1 and the bulge field population: the complex chemistry of Liller 1 closely resembles that of the bulge field (and of Terzan~5; see \citealp{origlia+25}), with a dominant old, sub-solar and $\alpha$-enhanced sub-population, and a younger, super-solar component with about solar-scaled $\alpha$-elements.

Hence, the comprehensive set of chemical abundances and patterns of iron-peak, neutron-capture, $\alpha$, and several other light elements measured in this study represents the high-resolution spectroscopic proof of the complex star formation and chemical enrichment history of Liller~1, indicating that it hosts multi-age and multi-metallicity sub-populations with remarkable chemical similarity to the bulge field. This strongly points to an in-situ formation and evolution, likely characterized by self-enrichment.

\begin{sidewaystable*}
\centering
\caption{Chemical abundances and corresponding errors of the observed metal-poor stars in Liller~1.  }
\label{tab3}
\scriptsize
\setlength{\tabcolsep}{2.0pt}
\renewcommand{\arraystretch}{1.2}
    \begin{tabular}{|c|c|c|c|c|c|c|c|c|c|c|c|c|c|c|c|c|c|c|c|c|c|}
    \hline\hline
    &&&&&&&&&&&&&&&&&&&&&\\
ID     & [Fe/H] & [C/H]  & [N/H]  & [O/H]  & [Na/H] & [Mg/H] & [Al/H] & [Si/H] & [S/H]  & [Ca/H] & [Sc/H] & [Ti/H] & [V/H] & [Cr/H] & [Co/H] & [Ni/H] & [Cu/H] & [Zn/H] & [Y/H] & [Nd/H] & $^{12}\mathrm{C}$/$^{13}\mathrm{C}$ \\
&&&&&&&&&&&&&&&&&&&&&\\
\hline \hline
100157 &-0.44   &-0.68   &+0.22   &+0.01   &-0.11   &+0.00   &+0.01   &-0.17   &+0.01   &-0.12   &+0.02   &-0.07   &-0.44 &- &- &- &- &- &- &-	&7.9\\
       &0.05(8) &0.08(3) &0.06(16)&0.10(1) &0.10(2) &0.10(1) &0.10(1) &0.07(2) &0.01(2) &0.10(2) &0.10(2) &0.02(2) &0.10(1) & & &	& & & & &\\\hline
100571 &-0.49   & -0.6   &+0.10   &-0.16   &+0.00   &-0.15   &-0.18   &-0.14   &-0.05   &-0.04   &-0.12   &-0.08   &-0.36 &-&-	&-	&-	&- &-0.59 &- &15.5\\
       &0.04(6) &0.02(4) &0.02(9) &0.08(4) &0.10(1) &0.10(1) &0.10(1) &0.10(1) &0.10(1) &0.06(2) &0.06(2) &0.06(2) &0.10(1) & & &	& &    &0.10(1)& &\\\hline
100658 &-0.44   &-0.63   &+0.18   &-0.04   &+0.06   &-0.12   &-0.15   &-0.10   &+0.05   &-0.04   &-0.16   &+0.00   &-0.37 &- &- &- &- &- &-0.38 &- &17.4\\
       &0.04(11)&0.02(5) &0.02(19)&0.04(5) &0.10(1) &0.10(1) &0.10(1) &0.01(2) &0.05(2) &0.10(1) &0.02(2) &0.02(3) &0.10(1) & & & & &     &0.10(1) & &\\\hline
100689 &-0.36   &-0.68   &+0.22	  &+0.02   &+0.06   &-0.04   &+0.07	  &-0.12   &+0.02   &-0.07   &-0.15	  &+0.04   &-0.36 &- &- &-	&- &-&-0.34	&- &18.4\\
       &0.03(10)&0.01(3) &0.03(12)&0.05(4) &0.10(1) &0.10(1) &0.10(1) &0.05(2) &0.08(2) &0.10(1) &0.04(2) &0.02(3) &0.10(1) & & & & &     &0.10(1) & &\\\hline
100756 &-0.34   &-0.60	 &+0.26   &+0.06   &-0.02   &+0.00	 &+0.10   &+0.01   &+0.07   &-0.09	 &-0.20	  &-0.06   &-0.45 &- &- &-&- &-   &-0.20 &-	&16.8\\
       &0.04(9) &0.02(4) &0.04(8) &0.10(2) &0.06(2) &0.10(1) &0.10(1) &0.01(2)	&0.10(1) &0.10(1) &0.04(2) &0.02(3) &0.10(1) & & & & &     &0.10(1) & &\\\hline
100760 &-0.41   &-0.66   &+0.34   &-0.12   &-0.07   &-0.04   &-0.08   &-0.09   &+0.04   &-0.16   &-0.31   &-0.17   &-0.34   &-0.40   &-0.36   &-0.42   &-0.29   &- &-0.39   &-0.40   &22.1\\
       &0.01(17)&0.03(10)&0.01(28)&0.06(3) &0.01(3) &0.02(2) &0.01(2) &0.01(3) &0.09(4) &0.05(3) &0.01(2) &0.04(5) &0.03(2) &0.03(2) &0.10(2) &0.10(2) &0.10(2) &  &0.10(1) &0.10(1) &\\\hline
100987 &-0.41   &-0.52   &+0.16   &-0.09   &-0.05   &-0.14   &-0.18   &-0.03   &+0.05   &-0.15   &-0.24   &-0.12   &-0.30 &- &- &- &- &- &-0.32 &- &18.3\\
       &0.02(6) &0.03(6) &0.04(9) &0.03(2) &0.10(1) &0.10(1) &0.10(1) &0.03(3)	&0.09(3) &0.01(2) &0.02(2) &0.04(3) &0.10(1) & & & & &    &0.10(1)	& &\\\hline
200119 &-0.45   &-0.95   &+0.35   &-0.19   &+0.08   &-0.13   &-0.12   &-0.17   &-0.05   &-0.14   &-0.07   &-0.07   &-0.39 &- &- &- &-&- &-0.49 &- &10.7\\
       &0.02(6) &0.04(4) &0.03(12)&0.05(4) &0.10(1) &0.10(1) &0.10(1) &0.09(4) &0.10(1) &0.10(1) &0.01(2) &0.10(1) &0.10(1) & & & & &    &0.10(1) & &\\\hline
300094 &-0.44   &-0.58   &+0.21   &+0.01   &-0.12   &+0.01   &-0.15   &-0.15   &-0.03   &-0.06   &-	      &-0.10   &-0.51   &-0.32   &-0.19   &-0.33   &-0.26   &-0.03   &- &-0.46	&11.6\\
       &0.02(7) &0.01(31)&0.05(6) &0.07(4) &0.02(2) &0.06(3) &0.10(1) &0.01(2) &0.10(1) &0.02(4) &	       &0.01(3) &0.03(4) &0.03(2) &0.10(1) &0.02(3) &0.01(2) &0.10(1) &  &0.10(1) &\\\hline
300097 &-0.45   &-1.06   &+0.40   &-0.15   &-0.05   &-0.04   &-0.05   &-0.03   &+0.06   &-0.07   &-0.26   &+0.05   &-0.47   &- &-	&- &- &- &-0.56 &- &10.3\\
       &0.07(7) &0.03(10)&0.03(13)&0.10(2) &0.10(1) &0.10(1) &0.10(1) &0.10(1) &0.10(1) &0.04(2) &0.06(2) &0.08(3) &0.10(1) & & & & &         &0.10(1) & &\\\hline
300162 &-0.40   &-0.64   &+0.23   &-0.11   &-0.19   &-0.02   &-0.09   &-0.06   &+0.08   &-0.04   &-0.15   &+0.00   &-0.47   &- &- &- &- &- &-0.38 &- &10.2\\
       &0.05(7) &0.03(4) &0.03(29)&0.06(2) &0.10(2) &0.10(1) &0.10(1) &0.05(2) &0.02(2) &0.02(2) &0.03(2) &0.01(2) &0.10(1) & & & & &      &0.10(1) & &\\\hline
300315 &-0.46   &-0.86   &+0.29   &-0.14   &-0.13   &-0.01   &-0.17   &-0.15   &+0.04   &-0.16   & -	  &-0.10   &-0.44   &-0.37   &-0.33   &-0.40   &-0.26   &-0.13   &- &-0.62   &8.8\\
       &0.03(7) &0.01(29)&0.04(9) &0.08(3) &0.05(2) &0.03(3) &0.03(2) &0.03(5) &0.10(1) &0.03(3) &        &0.04(2) &0.02(4) &0.04(3) &0.10(1) &0.02(2) &0.10(1) &0.10(1) &  &0.10(1) &\\\hline
300553 &-0.42   &-0.47   &+0.01   &-0.06   &-0.17   &-0.05   &-0.16   &-0.17   &-0.12   &-0.16   &-0.13   &-0.07   &-0.31   &-0.30   &-0.31   &-0.37   &-0.23   &-     &-0.67   &-0.36   &22.1\\
       &0.01(18)&0.01(36)&0.01(23)&0.07(3) &0.05(3) &0.02(3) &0.01(2) &0.01(5) &0.02(3) &0.02(3) &0.01(2) &0.02(5) &0.06(2) &0.02(2) &0.10(2) &0.07(2) &0.01(2) &      &0.10(1) &0.10(1) &\\\hline
300682 &-0.27   &-0.34   &+0.29   &-0.02   &+0.21   &+0.03   &+0.12   &+0.06   &+0.03   &+0.01   &+0.03   &+0.11   &-0.12   &-0.10   &-0.08   &-0.14   &-0.06   &- &-0.29   &-0.28   &26.7\\
       &0.01(19)&0.01(34)&0.01(40)&0.01(9) &0.02(4) &0.02(3) &0.01(2) &0.02(3) &0.05(4) &0.02(3) &0.04(2) &0.02(3) &0.01(2) &0.10(2) &0.10(2) &0.03(3) &0.08(2) &  &0.10(1) &0.10(1) &\\\hline
300701 &-0.37   &-0.55   &+0.28   &-0.03   &-0.19   &-0.12   &+0.09   &-0.10    &+0.13   &+0.05   &-0.16   &-0.01   &-0.42   &-	&- &- &-&- &-0.40 &- &25.2\\
       &0.04(8) &0.03(3) &0.03(24)&0.10(2) &0.10(1) &0.10(1) &0.10(1) &0.08(3) &0.10(1) &0.03(2) &0.04(2) &0.06(2) &0.10(1) & & & & &      &0.10(1) & &\\\hline
400519 &-0.40   &-0.69   &+0.18   &-0.14   &+0.04   &-0.06   &+0.01   &-0.06   &+0.03   &-0.02   &-0.19   &+0.04   &-0.41   &- &-	&-&- &- &-0.37   &-&14.7\\
       &0.06(6) &0.05(5) &0.04(9) &0.06(3)	&0.10(1) &0.10(1) &0.10(1) &0.01(2)	&0.05(3) &0.01(2) &0.01(2) &0.04(3) &0.10(1) & & & & &       &0.10(1) &  &\\\hline
400733 &-0.44   &-0.56   &+0.19   &-0.01   &-0.14   &+0.01   &-0.13   &-0.24   &-0.01   &-0.14   &-0.07   &-0.06   &-0.32   &- &- &-&- &- &-0.41 &- &20.3\\
       &0.03(7) &0.05(2) &0.02(9) &0.03(3)	&0.10(1) &0.10(1) &0.10(1) &0.10(2)	&0.10(2) &0.02(2) &0.05(2) &0.02(2) &0.10(1) & & & & &      &0.10(1) & &	\\\hline
400860 &-0.46   &-1.11   &+0.24   &-0.08   &+0.05   &-0.02   &+0.00   &-0.10   &-0.14   &-0.14   &-0.30   &-0.02   &-0.29 &- &- &-&- &- &- &- &16.1\\
       &0.03(6) &0.02(3) &0.04(8) &0.06(3)	&0.10(1) &0.10(1) &0.10(1) &0.06(2)	&0.10(1) &0.10(1) &0.05(2) &0.01(3) &0.10(1) & & & & & & & &\\
\hline\hline
\end{tabular}
\vspace{0.25cm}\\
Notes: Solar reference abundances are from \citealt{magg_22}. The quoted errors are the standard deviations divided by the square root of the number of lines used, indicated in brackets.
\end{sidewaystable*}

\begin{sidewaystable*}
\centering
\caption{Chemical abundances and corresponding errors of the observed metal-intermediate and metal-rich stars in Liller~1. }
\label{tab4}
\scriptsize
\setlength{\tabcolsep}{2.0pt}
\renewcommand{\arraystretch}{1.2}
    \begin{tabular}{|c|c|c|c|c|c|c|c|c|c|c|c|c|c|c|c|c|c|c|c|c|c|}
    \hline\hline
        &&&&&&&&&&&&&&&&&&&&&\\
ID     & [Fe/H] & [C/H]  & [N/H]  & [O/H]  & [Na/H] & [Mg/H] & [Al/H] & [Si/H] & [S/H]  & [Ca/H] & [Sc/H] & [Ti/H] & [V/H] & [Cr/H] & [Co/H] & [Ni/H] & [Cu/H] & [Zn/H] & [Y/H] & [Nd/H] & $^{12}\mathrm{C}$/$^{13}\mathrm{C}$ \\
    &&&&&&&&&&&&&&&&&&&&&\\
\hline\hline
100901 &-0.12   &-0.24   &+0.60   &+0.04   &+0.37   &+0.10   &+0.09   &+0.06   &+0.14   &+0.11   &+0.22   &+0.18   &+0.02   &- &- &- &- &- &-0.22 &- &21.2\\
       &0.04(11)&0.03(8) &0.01(26)&0.02(2) &0.10(1) &0.10(1) &0.10(1) &0.04(4) &0.05(4) &0.04(2) &0.02(2) &0.02(2) &0.10(1) & & & & &      &0.10(1)	& &\\\hline
400476 &-0.16   &-0.30   &+0.39   &+0.27   &+0.12   &+0.20   &+0.32   &-0.01   &+0.08   &-0.09   &-0.18   &+0.03   &-0.34   &- &- &- &- &- &-0.02 &- &21.1\\
       &0.07(6) &0.04(6) &0.03(14)&0.02(3) &0.10(1) &0.10(1) &0.10(1) &0.10(1) &0.10(1) &0.05(2) &0.05(2) &0.04(3) &0.10(1) & & & & &       &0.10(1) & &\\\hline
400778 &-0.17   &-0.44   &+0.60   &+0.03   &+0.27   &+0.04   &+0.06   &+0.05   &+0.20   &+0.07   &+0.03   &+0.09   &-0.02   &-0.20   &-0.10   &-0.16   &+0.01   &- &-0.16   &-0.18  &17.6\\ 
       &0.01(16)&0.01(24)&0.02(12)&0.02(5) &0.06(3) &0.03(3) &0.02(4) &0.05(3) &0.07(5) &0.04(5) &0.01(2) &0.01(4) &0.01(2) &0.06(2) &0.10(2) &0.07(3) &0.10(1) &  &0.10(1) &0.10(1) &\\\hline
400829 &-0.11   &-0.33   &+0.54   &+0.07   &+0.40   &+0.03   &+0.08   &+0.08   &+0.10   &+0.08   &+0.08   &+0.15   &+0.04   &-0.06   &-0.01   &-0.02   &+0.07   &- &-0.17   &-0.23 &24.2\\
       &0.01(33)&0.01(14)&0.01(14)&0.04(6) &0.05(4) &0.10(2) &0.02(3) &0.01(6) &0.03(5) &0.02(3) &0.10(1) &0.03(3) &0.01(2) &0.09(2) &0.10(2) &0.06(3) &0.10(1) &  &0.10(1) &0.10(1) &\\\hline
100437 &+0.21   &-0.30   &+1.01   &+0.18   &+0.62   &+0.25   &+0.31   &+0.22   &+0.20   &+0.19   &0.09    &+0.21   &-0.02   &+0.16   &+0.37   &+0.22   &+0.32   &- &+0.10   &-0.09  &13.2\\
       &0.01(19)&0.01(24)&0.01(34)&0.06(5) &0.02(3) &0.03(4) &0.02(3) &0.03(3) &0.06(4) &0.04(3) &0.04(2) &0.04(6) &0.01(2) &0.01(2) &0.10(1) &0.07(4) &0.10(1) &  &0.10(1) &0.10(1) &\\\hline
200179 &+0.28   &-0.01   &+1.00   &+0.28   &+0.80   &+0.34   &+0.31   &+0.24   &+0.30   &+0.38   &+0.80   &+0.48   &+0.23   &- &- &-	&- &- &0.31	&- &14.3\\
       &0.05(10)&0.10(1) &0.03(31)&0.02(2) &0.10(1) &0.10(1) &0.10(1) &0.02(2) &0.03(3) &0.02(2) &0.10(1) &0.10(1) &0.10(1) & & & & &         &0.10(1)	& &\\\hline
300614 &+0.27   &-0.08   &+0.89	  &+0.20   &+0.59   &+0.37   &+0.24   &+0.29   &+0.27   &+0.26   &+0.28   &+0.25   &+0.10   &+0.26   &+0.33   &+0.27   &+0.35   &- &+0.04   &+0.08  &18.8\\
       &0.01(19)&0.01(30)&0.02(39)&0.03(8) &0.02(4) &0.02(3) &0.01(2) &0.02(5) &0.10(5) &0.01(4) &0.10(1) &0.02(4) &0.05(2) &0.10(1) &0.10(1) &0.07(2) &0.10(1) &  &0.10(1) &0.10(1) &\\\hline
300727 &+0.28   &+0.18   &+0.83   &+0.40   &+0.63   &+0.23   &+0.38   &+0.28   &+0.28   &+0.31   &+0.49   &+0.53   &+0.29   &+0.18   &+0.35   &+0.39   &+0.39   &- &+0.27   &+0.29   &16.5\\
       &0.01(20)&0.01(23)&0.01(37)&0.02(4) &0.03(3) &0.01(2) &0.01(2) &0.01(7) &0.03(5) &0.01(3) &0.10(1) &0.02(3) &0.03(2) &0.10(2) &0.09(2) &0.01(2) &0.04(2) &  &0.10(1) &0.10(1) &\\\hline
387099 &+0.20   &-0.05   &+0.64   &+0.20   &+0.55   &+0.27   &+0.20   &+0.27   &+0.22   &+0.30   &-       &+0.26   &+0.09   &+0.17   &+0.29   &+0.25   &+0.26   &+0.19   &- &+0.13	&10.0\\
       &0.04(8) &0.01(20)&0.02(10)&0.02(6) &0.06(2) &0.03(4) &0.01(2) &0.02(5) &0.10(1) &0.02(5) &	       &0.02(7) &0.10(1) &0.10(2) &0.10(1) &0.05(2) &0.01(2) &0.10(1) &  &0.10(1)	&\\\hline
400065 &+0.22   &+0.01   &+0.81   &+0.21   &+0.64   &+0.28   &+0.18   &+0.24   &+0.35   &+0.30   &-	      &+0.27   &+0.22   &+0.14   &+0.27   &+0.34   &+0.25   &+0.27   &- &+0.07	&13.3\\ 
       &0.02(8) &0.01(30)&0.04(10)&0.04(9) &0.10(1) &0.05(4) &0.02(2) &0.04(4) &0.10(1) &0.02(6) &        &0.01(5) &0.05(6) &0.10(2) &0.10(1) &0.01(2) &0.10(1) &0.10(1) &  &0.10(1) &\\\hline
400087 &+0.20   &-0.11   &+0.92   &+0.24   &+0.60   &+0.20   &+0.23   &+0.19   &+0.22   &+0.28   &+0.67   &+0.42   &+0.13   &- &-	&- &- &- &0.12 &- &11.1\\
       &0.02(8) &0.06(2) &0.03(30)&0.02(2) &0.03(2) &0.10(1) &0.10(1) &0.07(3) &0.02(2) &0.10(1) &0.10(1) &0.10(1) &0.10(1) & & & & &        &0.10(1)	& &\\	\hline
400887 &+0.26   &-0.08   &+0.99   &+0.23   &+0.62   &+0.26   &+0.30   &+0.23   &+0.24   &+0.25   &+0.34   &+0.29   &+0.18   &+0.24   &+0.44   &+0.32   &+0.45   &- &- &+0.22 &21.3\\ 
       &0.01(19)&0.01(27)&0.03(37)&0.03(7) &0.03(4) &0.03(3) &0.02(4) &0.02(4) &0.03(7) &0.02(4) &0.01(3) &0.02(5) &0.01(2) &0.05(3) &0.10(1) &0.06(2) &0.10(1) & &   &0.04(2) &\\
\hline\hline
\end{tabular}
\vspace{0.25cm}\\
Notes: Solar reference abundances are from \citealt{magg_22}. The quoted errors are the standard deviations divided by the square root of the number of lines used, indicated in brackets.
\end{sidewaystable*}

\begin{acknowledgements}
This work is part of the project {\it "GENESIS - Searching for the primordial structures of the Universe in the heart of the Galaxy"} (Advanced Grant FIS-2024-02056, PI:Ferraro), funded by the Italian MUR through the Fondo Italiano per la Scienza call. 
\end{acknowledgements}

\bibliographystyle{aasjournal}  
\bibliography{biblio}           

@ARTICLE{zullo+26,
       author = {{Zullo}, G. and {Pallanca}, C. and {Ferraro}, F.~R. and {Lanzoni}, B. and {Origlia}, L. and {Massari}, D. and {Dalessandro}, E. and {Fanelli}, C. and {Cadelano}, M. and {Vesperini}, E. and {Crociati}, C. and {Rich}, R.~M. and {Valenti}, E.},
        title = "{The multi-age stellar populations of Terzan 5 as revealed by JWST}",
      journal = {\aap},
         year = 2026,
        month = may,
       volume = {709},
          eid = {A212},
        pages = {A212},
          doi = {10.1051/0004-6361/202659349},
archivePrefix = {arXiv},
       eprint = {2604.00098},
 primaryClass = {astro-ph.GA},
       adsurl = {https://ui.adsabs.harvard.edu/abs/2026A&A...709A.212Z}
}

@ARTICLE{ferraro+26,
       author = {{Ferraro}, F.~R. and {Vesperini}, E. and {Lanzoni}, B. and {Romano}, D. and {Origlia}, L. and {Pallanca}, C. and {Fanelli}, C. and {Calura}, F. and {Dalessandro}, E. and {Massari}, D. and {Zullo}, G. and {Cadelano}, M.},
        title = "{Bulge fossil fragments as a new population of factories of gravitational wave sources in the Galaxy}",
      journal = {\aap},
         year = 2026,
        month = may,
       volume = {709},
          eid = {A163},
        pages = {A163},
          doi = {10.1051/0004-6361/202556993},
archivePrefix = {arXiv},
       eprint = {2603.25127},
 primaryClass = {astro-ph.GA},
       adsurl = {https://ui.adsabs.harvard.edu/abs/2026A&A...709A.163F}
}

@ARTICLE{gratton+07,
       author = {{Gratton}, R.~G. and {Lucatello}, S. and {Bragaglia}, A. and {Carretta}, E. and {Cassisi}, S. and {Momany}, Y. and {Pancino}, E. and {Valenti}, E. and {Caloi}, V. and {Claudi}, R. and {D'Antona}, F. and {Desidera}, S. and {Fran{\c{c}}ois}, P. and {James}, G. and {Moehler}, S. and {Ortolani}, S. and {Pasquini}, L. and {Piotto}, G. and {Recio-Blanco}, A.},
        title = "{Na-O anticorrelation and horizontal branches. V. The Na-O anticorrelation in <ASTROBJ>NGC 6441</ASTROBJ> from Giraffe spectra}",
      journal = {\aap},
         year = 2007,
        month = mar,
       volume = {464},
       number = {3},
        pages = {953-965},
          doi = {10.1051/0004-6361:20066061},
archivePrefix = {arXiv},
       eprint = {astro-ph/0701179},
 primaryClass = {astro-ph},
       adsurl = {https://ui.adsabs.harvard.edu/abs/2007A&A...464..953G}
}

@ARTICLE{carretta+23,
       author = {{Carretta}, Eugenio and {Bragaglia}, Angela},
        title = "{Chemistry of multiple stellar populations in the mono-metallic, in situ, bulge globular cluster NGC 6388}",
      journal = {\aap},
         year = 2023,
        month = sep,
       volume = {677},
          eid = {A73},
        pages = {A73},
          doi = {10.1051/0004-6361/202346174},
archivePrefix = {arXiv},
       eprint = {2307.05478},
 primaryClass = {astro-ph.GA},
       adsurl = {https://ui.adsabs.harvard.edu/abs/2023A&A...677A..73C}
}

@INPROCEEDINGS{renzini86,
       author = {{Renzini}, Alvio and {Buzzoni}, Alberto},
        title = "{Global properties of stellar populations and the spectral evolution of galaxies.}",
    booktitle = {Spectral Evolution of Galaxies},
         year = 1986,
       editor = {{Chiosi}, Cesare and {Renzini}, Alvio},
       series = {Astrophysics and Space Science Library},
       volume = {122},
        month = jan,
        pages = {195-231},
          doi = {10.1007/978-94-009-4598-2_19},
       adsurl = {https://ui.adsabs.harvard.edu/abs/1986ASSL..122..195R}
}

@ARTICLE{pietrinferni+13,
       author = {{Pietrinferni}, Adriano and {Cassisi}, Santi and {Salaris}, Maurizio and {Hidalgo}, Sebastian},
        title = "{The BaSTI Stellar Evolution Database: models for extremely metal-poor and super-metal-rich stellar populations}",
      journal = {\aap},
         year = 2013,
        month = oct,
       volume = {558},
          eid = {A46},
        pages = {A46},
          doi = {10.1051/0004-6361/201321950},
archivePrefix = {arXiv},
       eprint = {1308.3850},
 primaryClass = {astro-ph.SR},
       adsurl = {https://ui.adsabs.harvard.edu/abs/2013A&A...558A..46P}
}

@ARTICLE{liptrott26,
       author = {{Liptrott}, Anna and {Schiavon}, Ricardo P. and {Mason}, Andrew C. and {Kamann}, Sebastian and {Anguiano}, Borja and {Cohen}, Roger E. and {Fern{\'a}ndez-Trincado}, Jos{\'e} G. and {Horta}, Danny and {Majewski}, Steven R. and {Minniti}, Dante and {Nataf}, David M. and {O'Connor}, Michael J.~W. and {Wearne}, Dominic},
        title = "{Is Liller 1 a building block of the Galactic bulge? -- Evidence with APOGEE}",
      journal = {arXiv e-prints},
         year = 2025,
        month = oct,
          eid = {arXiv:2510.07411},
        pages = {arXiv:2510.07411},
          doi = {10.48550/arXiv.2510.07411},
archivePrefix = {arXiv},
       eprint = {2510.07411},
 primaryClass = {astro-ph.GA},
       adsurl = {https://ui.adsabs.harvard.edu/abs/2025arXiv251007411L}
}

@ARTICLE{salessilva24,
       author = {{Sales-Silva}, J.~V. and {Cunha}, K. and {Smith}, V.~V. and {Daflon}, S. and {Souto}, D. and {Guer{\c{c}}o}, R. and {Queiroz}, A. and {Chiappini}, C. and {Hayes}, C.~R. and {Masseron}, T. and {Hasselquist}, Sten and {Horta}, D. and {Prantzos}, N. and {Zoccali}, M. and {Allende Prieto}, C. and {Barbuy}, B. and {Beaton}, R. and {Bizyaev}, D. and {Fern{\'a}ndez-Trincado}, J.~G. and {Frinchaboy}, P.~M. and {Holtzman}, J.~A. and {Johnson}, J.~A. and {J{\"o}nsson}, Henrik and {Majewski}, S.~R. and {Minniti}, D. and {Nidever}, D.~L. and {Schiavon}, R.~P. and {Schultheis}, M. and {Sobeck}, J. and {Stringfellow}, G.~S. and {Zasowski}, G.},
        title = "{A Perspective on the Milky Way Bulge Bar as Seen from the Neutron-capture Elements Cerium and Neodymium with APOGEE}",
      journal = {\apj},
         year = 2024,
        month = apr,
       volume = {965},
       number = {2},
          eid = {119},
        pages = {119},
          doi = {10.3847/1538-4357/ad28c2},
archivePrefix = {arXiv},
       eprint = {2402.14898},
 primaryClass = {astro-ph.GA},
       adsurl = {https://ui.adsabs.harvard.edu/abs/2024ApJ...965..119S}
}

@ARTICLE{swaelmen16,
       author = {{Van der Swaelmen}, M. and {Barbuy}, B. and {Hill}, V. and {Zoccali}, M. and {Minniti}, D. and {Ortolani}, S. and {G{\'o}mez}, A.},
        title = "{Heavy elements Ba, La, Ce, Nd, and Eu in 56 Galactic bulge red giants}",
      journal = {\aap},
         year = 2016,
        month = jan,
       volume = {586},
          eid = {A1},
        pages = {A1},
          doi = {10.1051/0004-6361/201525709},
archivePrefix = {arXiv},
       eprint = {1511.03919},
 primaryClass = {astro-ph.GA},
       adsurl = {https://ui.adsabs.harvard.edu/abs/2016A&A...586A...1V}
}

@ARTICLE{johnson12,
       author = {{Johnson}, Christian I. and {Rich}, R. Michael and {Kobayashi}, Chiaki and {Fulbright}, Jon P.},
        title = "{Constraints on the Formation of the Galactic Bulge from Na, Al, and Heavy-element Abundances in Plaut's Field}",
      journal = {\apj},
         year = 2012,
        month = apr,
       volume = {749},
       number = {2},
          eid = {175},
        pages = {175},
          doi = {10.1088/0004-637X/749/2/175},
archivePrefix = {arXiv},
       eprint = {1202.4481},
 primaryClass = {astro-ph.SR},
       adsurl = {https://ui.adsabs.harvard.edu/abs/2012ApJ...749..175J}
}

@ARTICLE{origlia+25,
       author = {{Origlia}, L. and {Ferraro}, F.~R. and {Fanelli}, C. and {Lanzoni}, B. and {Massari}, D. and {Dalessandro}, E. and {Pallanca}, C.},
        title = "{The manifest link between Terzan 5 and the Galactic bulge}",
      journal = {\aap},
         year = 2025,
        month = may,
       volume = {697},
          eid = {A19},
        pages = {A19},
          doi = {10.1051/0004-6361/202452110},
archivePrefix = {arXiv},
       eprint = {2503.17258},
 primaryClass = {astro-ph.GA},
       adsurl = {https://ui.adsabs.harvard.edu/abs/2025A&A...697A..19O}
}

@ARTICLE{ferraro+25a,
       author = {{Ferraro}, F.~R. and {Chiappino}, L. and {Bartolomei}, A. and {Origlia}, L. and {Fanelli}, C. and {Lanzoni}, B. and {Pallanca}, C. and {Loriga}, M. and {Leanza}, S. and {Valenti}, E. and {Romano}, D. and {Mucciarelli}, A. and {Massari}, D. and {Cadelano}, M. and {Dalessandro}, E. and {Crociati}, C. and {Rich}, R.~M.},
        title = "{The Bulge Cluster Origin (BulCO) survey at the ESO-VLT: Probing the early history of the Milky Way assembly. Design and first results in Liller 1}",
      journal = {\aap},
         year = 2025,
        month = apr,
       volume = {696},
          eid = {A179},
        pages = {A179},
          doi = {10.1051/0004-6361/202554092},
archivePrefix = {arXiv},
       eprint = {2503.14642},
 primaryClass = {astro-ph.GA},
       adsurl = {https://ui.adsabs.harvard.edu/abs/2025A&A...696A.179F}
}

@ARTICLE{dorn+23,
       author = {{Dorn}, R.~J. and {Bristow}, P. and {Smoker}, J.~V. and {Rodler}, F. and {Lavail}, A. and {Accardo}, M. and {van den Ancker}, M. and {Baade}, D. and {Baruffolo}, A. and {Courtney-Barrer}, B. and {Blanco}, L. and {Brucalassi}, A. and {Cumani}, C. and {Follert}, R. and {Haimerl}, A. and {Hatzes}, A. and {Haug}, M. and {Heiter}, U. and {Hinterschuster}, R. and {Hubin}, N. and {Ives}, D.~J. and {Jung}, Y. and {Jones}, M. and {Kaeufl}, H. -U. and {Kirchbauer}, J. -P. and {Klein}, B. and {Kochukhov}, O. and {Korhonen}, H.~H. and {K{\"o}hler}, J. and {Lizon}, J. -L. and {Moins}, C. and {Molina-Conde}, I. and {Marquart}, T. and {Neeser}, M. and {Oliva}, E. and {Pallanca}, L. and {Pasquini}, L. and {Paufique}, J. and {Piskunov}, N. and {Reiners}, A. and {Schneller}, D. and {Schmutzer}, R. and {Seemann}, U. and {Slumstrup}, D. and {Smette}, A. and {Stegmeier}, J. and {Stempels}, E. and {Tordo}, S. and {Valenti}, E. and {Valenzuela}, J.~J. and {Vernet}, J. and {Vinther}, J. and {Wehrhahn}, A.},
        title = "{CRIRES$^{+}$ on sky at the ESO Very Large Telescope. Observing the Universe at infrared wavelengths and high spectral resolution}",
      journal = {\aap},
         year = 2023,
        month = mar,
       volume = {671},
          eid = {A24},
        pages = {A24},
          doi = {10.1051/0004-6361/202245217},
archivePrefix = {arXiv},
       eprint = {2301.08048},
 primaryClass = {astro-ph.IM},
       adsurl = {https://ui.adsabs.harvard.edu/abs/2023A&A...671A..24D}
}

@ARTICLE{dorn+14,
       author = {{Dorn}, R.~J. and {Anglada-Escude}, G. and {Baade}, D. and {Bristow}, P. and {Follert}, R. and {Gojak}, D. and {Grunhut}, J. and {Hatzes}, A. and {Heiter}, U. and {Hilker}, M. and {Ives}, D.~J. and {Jung}, Y. and {K{\"a}ufl}, H. -U. and {Kerber}, F. and {Klein}, B. and {Lizon}, J. -L. and {Lockhart}, M. and {L{\"o}winger}, T. and {Marquart}, T. and {Oliva}, E. and {Origlia}, L. and {Pasquini}, L. and {Paufique}, J. and {Piskunov}, N. and {Pozna}, E. and {Reiners}, A. and {Smette}, A. and {Smoker}, J. and {Seemann}, U. and {Stempels}, E. and {Valenti}, E.},
        title = "{CRIRES+: Exploring the Cold Universe at High Spectral Resolution}",
      journal = {The Messenger},
         year = 2014,
        month = jun,
       volume = {156},
        pages = {7-11},
       adsurl = {https://ui.adsabs.harvard.edu/abs/2014Msngr.156....7D}
}

@INPROCEEDINGS{kaufl+04,
       author = {{Kaeufl}, Hans-Ulrich and {Ballester}, Pascal and {Biereichel}, Peter and {Delabre}, Bernard and {Donaldson}, Rob and {Dorn}, Reinhold and {Fedrigo}, Enrico and {Finger}, Gert and {Fischer}, Gerhard and {Franza}, Francis and {Gojak}, Domingo and {Huster}, Gotthard and {Jung}, Yves and {Lizon}, Jean-Louis and {Mehrgan}, Leander and {Meyer}, Manfred and {Moorwood}, Alan and {Pirard}, Jean-Francois and {Paufique}, Jerome and {Pozna}, Esther and {Siebenmorgen}, Ralf and {Silber}, Armin and {Stegmeier}, Joerg and {Wegerer}, Stefan},
        title = "{CRIRES: a high-resolution infrared spectrograph for ESO's VLT}",
    booktitle = {Ground-based Instrumentation for Astronomy},
         year = 2004,
       editor = {{Moorwood}, Alan F.~M. and {Iye}, Masanori},
       series = {Society of Photo-Optical Instrumentation Engineers (SPIE) Conference Series},
       volume = {5492},
        month = sep,
        pages = {1218-1227},
          doi = {10.1117/12.551480},
       adsurl = {https://ui.adsabs.harvard.edu/abs/2004SPIE.5492.1218K}
}

@ARTICLE{marigo+17,
       author = {{Marigo}, Paola and {Girardi}, L{\'e}o and {Bressan}, Alessandro and {Rosenfield}, Philip and {Aringer}, Bernhard and {Chen}, Yang and {Dussin}, Marco and {Nanni}, Ambra and {Pastorelli}, Giada and {Rodrigues}, Tha{\'\i}se S. and {Trabucchi}, Michele and {Bladh}, Sara and {Dalcanton}, Julianne and {Groenewegen}, Martin A.~T. and {Montalb{\'a}n}, Josefina and {Wood}, Peter R.},
        title = "{A New Generation of PARSEC-COLIBRI Stellar Isochrones Including the TP-AGB Phase}",
      journal = {\apj},
         year = 2017,
        month = jan,
       volume = {835},
       number = {1},
          eid = {77},
        pages = {77},
          doi = {10.3847/1538-4357/835/1/77},
archivePrefix = {arXiv},
       eprint = {1701.08510},
 primaryClass = {astro-ph.SR},
       adsurl = {https://ui.adsabs.harvard.edu/abs/2017ApJ...835...77M}
}

@ARTICLE{crociati+24,
       author = {{Crociati}, C. and {Cignoni}, M. and {Dalessandro}, E. and {Pallanca}, C. and {Massari}, D. and {Ferraro}, F.~R. and {Lanzoni}, B. and {Origlia}, L. and {Valenti}, E.},
        title = "{The star formation history of the first bulge fossil fragment candidate Terzan 5}",
      journal = {\aap},
         year = 2024,
        month = nov,
       volume = {691},
          eid = {A311},
        pages = {A311},
          doi = {10.1051/0004-6361/202451174},
archivePrefix = {arXiv},
       eprint = {2410.16971},
 primaryClass = {astro-ph.GA},
       adsurl = {https://ui.adsabs.harvard.edu/abs/2024A&A...691A.311C}
}

@ARTICLE{fanelli+24,
       author = {{Fanelli}, C. and {Origlia}, L. and {Rich}, R.~M. and {Ferraro}, F.~R. and {Alvarez Garay}, D.~A. and {Chiappino}, L. and {Lanzoni}, B. and {Pallanca}, C. and {Crociati}, C. and {Dalessandro}, E.},
        title = "{Multi-iron subpopulations in Liller 1 from high-resolution H-band spectroscopy}",
      journal = {\aap},
         year = 2024,
        month = oct,
       volume = {690},
          eid = {A139},
        pages = {A139},
          doi = {10.1051/0004-6361/202451030},
archivePrefix = {arXiv},
       eprint = {2408.12649},
 primaryClass = {astro-ph.GA},
       adsurl = {https://ui.adsabs.harvard.edu/abs/2024A&A...690A.139F}
}

@ARTICLE{bournaud09,
       author = {{Bournaud}, Fr{\'e}d{\'e}ric and {Elmegreen}, Bruce G.},
        title = "{Unstable Disks at High Redshift: Evidence for Smooth Accretion in Galaxy Formation}",
      journal = {\apjl},
         year = 2009,
        month = apr,
       volume = {694},
       number = {2},
        pages = {L158-L161},
          doi = {10.1088/0004-637X/694/2/L158},
archivePrefix = {arXiv},
       eprint = {0902.2806},
 primaryClass = {astro-ph.CO},
       adsurl = {https://ui.adsabs.harvard.edu/abs/2009ApJ...694L.158B}
}

@INPROCEEDINGS{bournaud16,
       author = {{Bournaud}, Fr{\'e}d{\'e}ric},
        title = "{Bulge Growth Through Disc Instabilities in High-Redshift Galaxies}",
    booktitle = {Galactic Bulges},
         year = 2016,
       editor = {{Laurikainen}, Eija and {Peletier}, Reynier and {Gadotti}, Dimitri},
       series = {Astrophysics and Space Science Library},
       volume = {418},
        month = jan,
        pages = {355},
          doi = {10.1007/978-3-319-19378-6_13},
archivePrefix = {arXiv},
       eprint = {1503.07660},
 primaryClass = {astro-ph.GA},
       adsurl = {https://ui.adsabs.harvard.edu/abs/2016ASSL..418..355B}
}

@ARTICLE{elme09,
       author = {{Elmegreen}, Bruce G. and {Elmegreen}, Debra Meloy and {Fernandez}, Maria Ximena and {Lemonias}, Jenna Jo},
        title = "{Bulge and Clump Evolution in Hubble Ultra Deep Field Clump Clusters, Chains and Spiral Galaxies}",
      journal = {\apj},
         year = 2009,
        month = feb,
       volume = {692},
       number = {1},
        pages = {12-31},
          doi = {10.1088/0004-637X/692/1/12},
archivePrefix = {arXiv},
       eprint = {0810.5404},
 primaryClass = {astro-ph},
       adsurl = {https://ui.adsabs.harvard.edu/abs/2009ApJ...692...12E}
}

@ARTICLE{genzel11,
       author = {{Genzel}, R. and {Newman}, S. and {Jones}, T. and {F{\"o}rster Schreiber}, N.~M. and {Shapiro}, K. and {Genel}, S. and {Lilly}, S.~J. and {Renzini}, A. and {Tacconi}, L.~J. and {Bouch{\'e}}, N. and {Burkert}, A. and {Cresci}, G. and {Buschkamp}, P. and {Carollo}, C.~M. and {Ceverino}, D. and {Davies}, R. and {Dekel}, A. and {Eisenhauer}, F. and {Hicks}, E. and {Kurk}, J. and {Lutz}, D. and {Mancini}, C. and {Naab}, T. and {Peng}, Y. and {Sternberg}, A. and {Vergani}, D. and {Zamorani}, G.},
        title = "{The Sins Survey of z \raisebox{-0.5ex}\textasciitilde 2 Galaxy Kinematics: Properties of the Giant Star-forming Clumps}",
      journal = {\apj},
         year = 2011,
        month = jun,
       volume = {733},
       number = {2},
          eid = {101},
        pages = {101},
          doi = {10.1088/0004-637X/733/2/101},
archivePrefix = {arXiv},
       eprint = {1011.5360},
 primaryClass = {astro-ph.CO},
       adsurl = {https://ui.adsabs.harvard.edu/abs/2011ApJ...733..101G}
}

@ARTICLE{tacchella15,
       author = {{Tacchella}, S. and {Lang}, P. and {Carollo}, C.~M. and {F{\"o}rster Schreiber}, N.~M. and {Renzini}, A. and {Shapley}, A.~E. and {Wuyts}, S. and {Cresci}, G. and {Genzel}, R. and {Lilly}, S.~J. and {Mancini}, C. and {Newman}, S.~F. and {Tacconi}, L.~J. and {Zamorani}, G. and {Davies}, R.~I. and {Kurk}, J. and {Pozzetti}, L.},
        title = "{SINS/zC-SINF Survey of z {\ensuremath{\sim}} 2 Galaxy Kinematics: Rest-frame Morphology, Structure, and Colors from Near-infrared Hubble Space Telescope Imaging}",
      journal = {\apj},
         year = 2015,
        month = apr,
       volume = {802},
       number = {2},
          eid = {101},
        pages = {101},
          doi = {10.1088/0004-637X/802/2/101},
archivePrefix = {arXiv},
       eprint = {1411.7034},
 primaryClass = {astro-ph.GA},
       adsurl = {https://ui.adsabs.harvard.edu/abs/2015ApJ...802..101T}
}

@ARTICLE{elme08,
       author = {{Elmegreen}, Bruce G. and {Bournaud}, Fr{\'e}d{\'e}ric and {Elmegreen}, Debra Meloy},
        title = "{Bulge Formation by the Coalescence of Giant Clumps in Primordial Disk Galaxies}",
      journal = {\apj},
         year = 2008,
        month = nov,
       volume = {688},
       number = {1},
        pages = {67-77},
          doi = {10.1086/592190},
archivePrefix = {arXiv},
       eprint = {0808.0716},
 primaryClass = {astro-ph},
       adsurl = {https://ui.adsabs.harvard.edu/abs/2008ApJ...688...67E}
}

@ARTICLE{pfeffer21,
       author = {{Pfeffer}, Joel and {Lardo}, Carmela and {Bastian}, Nate and {Saracino}, Sara and {Kamann}, Sebastian},
        title = "{The accreted nuclear clusters of the Milky Way}",
      journal = {\mnras},
         year = 2021,
        month = jan,
       volume = {500},
       number = {2},
        pages = {2514-2524},
          doi = {10.1093/mnras/staa3407},
archivePrefix = {arXiv},
       eprint = {2011.02042},
 primaryClass = {astro-ph.GA},
       adsurl = {https://ui.adsabs.harvard.edu/abs/2021MNRAS.500.2514P}
}

@ARTICLE{mckenzie18,
       author = {{McKenzie}, M. and {Bekki}, K.},
        title = "{A new model for the multiple stellar populations within Terzan 5}",
      journal = {\mnras},
         year = 2018,
        month = sep,
       volume = {479},
       number = {3},
        pages = {3126-3141},
          doi = {10.1093/mnras/sty1557},
archivePrefix = {arXiv},
       eprint = {1806.04824},
 primaryClass = {astro-ph.GA},
       adsurl = {https://ui.adsabs.harvard.edu/abs/2018MNRAS.479.3126M}
}

@ARTICLE{valenti10,
       author = {{Valenti}, E. and {Ferraro}, F.~R. and {Origlia}, L.},
        title = "{Near-infrared properties of 12 globular clusters towards the inner bulge of the Galaxy}",
      journal = {\mnras},
         year = 2010,
        month = mar,
       volume = {402},
       number = {3},
        pages = {1729-1739},
          doi = {10.1111/j.1365-2966.2009.15991.x},
archivePrefix = {arXiv},
       eprint = {0911.1264},
 primaryClass = {astro-ph.GA},
       adsurl = {https://ui.adsabs.harvard.edu/abs/2010MNRAS.402.1729V}
}

@ARTICLE{fanelli21,
       author = {{Fanelli}, C. and {Origlia}, L. and {Oliva}, E. and {Mucciarelli}, A. and {Sanna}, N. and {Dalessandro}, E. and {Romano}, D.},
        title = "{Stellar population astrophysics (SPA) with the TNG. The Arcturus Lab}",
      journal = {\aap},
         year = 2021,
        month = jan,
       volume = {645},
          eid = {A19},
        pages = {A19},
          doi = {10.1051/0004-6361/202039397},
archivePrefix = {arXiv},
       eprint = {2011.12321},
 primaryClass = {astro-ph.SR},
       adsurl = {https://ui.adsabs.harvard.edu/abs/2021A&A...645A..19F}
}

@ARTICLE{rich12,
       author = {{Rich}, R.~M. and {Origlia}, L. and {Valenti}, E.},
        title = "{Detailed Abundances for M Giants in Two Inner Bulge Fields from Infrared Spectroscopy}",
      journal = {\apj},
         year = 2012,
        month = feb,
       volume = {746},
       number = {1},
          eid = {59},
        pages = {59},
          doi = {10.1088/0004-637X/746/1/59},
archivePrefix = {arXiv},
       eprint = {1112.0306},
 primaryClass = {astro-ph.GA},
       adsurl = {https://ui.adsabs.harvard.edu/abs/2012ApJ...746...59R}
}

@ARTICLE{carretta09,
       author = {{Carretta}, E. and {Bragaglia}, A. and {Gratton}, R.~G. and {Lucatello}, S. and {Catanzaro}, G. and {Leone}, F. and {Bellazzini}, M. and {Claudi}, R. and {D'Orazi}, V. and {Momany}, Y. and {Ortolani}, S. and {Pancino}, E. and {Piotto}, G. and {Recio-Blanco}, A. and {Sabbi}, E.},
        title = "{Na-O anticorrelation and HB. VII. The chemical composition of first and second-generation stars in 15 globular clusters from GIRAFFE spectra}",
      journal = {\aap},
         year = 2009,
        month = oct,
       volume = {505},
       number = {1},
        pages = {117-138},
          doi = {10.1051/0004-6361/200912096},
archivePrefix = {arXiv},
       eprint = {0909.2938},
 primaryClass = {astro-ph.GA},
       adsurl = {https://ui.adsabs.harvard.edu/abs/2009A&A...505..117C}
}

@ARTICLE{romano23,
       author = {{Romano}, Donatella and {Ferraro}, Francesco R. and {Origlia}, Livia and {Zwart}, Simon Portegies and {Lanzoni}, Barbara and {Crociati}, Chiara and {Massari}, Davide and {Dalessandro}, Emanuele and {Mucciarelli}, Alessio and {Rich}, R. Michael and {Calura}, Francesco and {Matteucci}, Francesca},
        title = "{Modeling the Chemical Enrichment History of the Bulge Fossil Fragment Terzan 5}",
      journal = {\apj},
         year = 2023,
        month = jul,
       volume = {951},
       number = {2},
          eid = {85},
        pages = {85},
          doi = {10.3847/1538-4357/acd8ba},
archivePrefix = {arXiv},
       eprint = {2305.15355},
 primaryClass = {astro-ph.GA},
       adsurl = {https://ui.adsabs.harvard.edu/abs/2023ApJ...951...85R}
}

@ARTICLE{johnson_14,
       author = {{Johnson}, Christian I. and {Rich}, R. Michael and {Kobayashi}, Chiaki and {Kunder}, Andrea and {Koch}, Andreas},
        title = "{Light, Alpha, and Fe-peak Element Abundances in the Galactic Bulge}",
      journal = {\aj},
         year = 2014,
        month = oct,
       volume = {148},
       number = {4},
          eid = {67},
        pages = {67},
          doi = {10.1088/0004-6256/148/4/67},
archivePrefix = {arXiv},
       eprint = {1407.2282},
 primaryClass = {astro-ph.SR},
       adsurl = {https://ui.adsabs.harvard.edu/abs/2014AJ....148...67J}
}

@ARTICLE{bressan_12,
 author = {{Bressan}, Alessandro and {Marigo}, Paola and {Girardi}, L{\'e}o. and {Salasnich}, Bernardo and {Dal Cero}, Claudia and {Rubele}, Stefano and {Nanni}, Ambra},
        title = "{PARSEC: stellar tracks and isochrones with the PAdova and TRieste Stellar Evolution Code}",
      journal = {\mnras},
         year = 2012,
        month = nov,
       volume = {427},
       number = {1},
        pages = {127-145},
          doi = {10.1111/j.1365-2966.2012.21948.x},
archivePrefix = {arXiv},
       eprint = {1208.4498},
 primaryClass = {astro-ph.SR},
       adsurl = {https://ui.adsabs.harvard.edu/abs/2012MNRAS.427..127B}
}

@ARTICLE{Ferraro_09,
       author = {{Ferraro}, F.~R. and {Dalessandro}, E. and {Mucciarelli}, A. and {Beccari}, G. and {Rich}, R.~M. and {Origlia}, L. and {Lanzoni}, B. and {Rood}, R.~T. and {Valenti}, E. and {Bellazzini}, M. and {Ransom}, S.~M. and {Cocozza}, G.},
        title = "{The cluster Terzan 5 as a remnant of a primordial building block of the Galactic bulge}",
      journal = {\nat},
         year = 2009,
        month = nov,
       volume = {462},
       number = {7272},
        pages = {483-486},
          doi = {10.1038/nature08581},
archivePrefix = {arXiv},
       eprint = {0912.0192},
 primaryClass = {astro-ph.GA},
       adsurl = {https://ui.adsabs.harvard.edu/abs/2009Natur.462..483F}
}

@ARTICLE{Origlia_11,
       author = {{Origlia}, L. and {Rich}, R.~M. and {Ferraro}, F.~R. and {Lanzoni}, B. and {Bellazzini}, M. and {Dalessandro}, E. and {Mucciarelli}, A. and {Valenti}, E. and {Beccari}, G.},
        title = "{Spectroscopy Unveils the Complex Nature of Terzan 5}",
      journal = {\apjl},
         year = 2011,
        month = jan,
       volume = {726},
       number = {2},
          eid = {L20},
        pages = {L20},
          doi = {10.1088/2041-8205/726/2/L20},
archivePrefix = {arXiv},
       eprint = {1012.2047},
 primaryClass = {astro-ph.GA},
       adsurl = {https://ui.adsabs.harvard.edu/abs/2011ApJ...726L..20O}
}

@ARTICLE{Origlia_13,
       author = {{Origlia}, L. and {Massari}, D. and {Rich}, R.~M. and {Mucciarelli}, A. and {Ferraro}, F.~R. and {Dalessandro}, E. and {Lanzoni}, B.},
        title = "{The Terzan 5 Puzzle: Discovery of a Third, Metal-poor Component}",
      journal = {\apjl},
         year = 2013,
        month = dec,
       volume = {779},
       number = {1},
          eid = {L5},
        pages = {L5},
          doi = {10.1088/2041-8205/779/1/L5},
archivePrefix = {arXiv},
       eprint = {1311.1706},
 primaryClass = {astro-ph.GA},
       adsurl = {https://ui.adsabs.harvard.edu/abs/2013ApJ...779L...5O}
}

@ARTICLE{Massari_14,
       author = {{Massari}, D. and {Mucciarelli}, A. and {Ferraro}, F.~R. and {Origlia}, L. and {Rich}, R.~M. and {Lanzoni}, B. and {Dalessandro}, E. and {Valenti}, E. and {Ibata}, R. and {Lovisi}, L. and {Bellazzini}, M. and {Reitzel}, D.},
        title = "{Ceci N'est Pas a Globular Cluster: The Metallicity Distribution of the Stellar System Terzan 5}",
      journal = {\apj},
         year = 2014,
        month = nov,
       volume = {795},
       number = {1},
          eid = {22},
        pages = {22},
          doi = {10.1088/0004-637X/795/1/22},
archivePrefix = {arXiv},
       eprint = {1409.1682},
 primaryClass = {astro-ph.SR},
       adsurl = {https://ui.adsabs.harvard.edu/abs/2014ApJ...795...22M}
}

@ARTICLE{saracino_15,
       author = {{Saracino}, S. and {Dalessandro}, E. and {Ferraro}, F.~R. and {Lanzoni}, B. and {Geisler}, D. and {Mauro}, F. and {Villanova}, S. and {Moni Bidin}, C. and {Miocchi}, P. and {Massari}, D.},
        title = "{GEMINI/GeMS Observations Unveil the Structure of the Heavily Obscured Globular Cluster Liller 1.}",
      journal = {\apj},
         year = 2015,
        month = jun,
       volume = {806},
       number = {2},
          eid = {152},
        pages = {152},
          doi = {10.1088/0004-637X/806/2/152},
archivePrefix = {arXiv},
       eprint = {1505.00568},
 primaryClass = {astro-ph.SR},
       adsurl = {https://ui.adsabs.harvard.edu/abs/2015ApJ...806..152S}
}

@ARTICLE{magg_22,
       author = {{Magg}, Ekaterina and {Bergemann}, Maria and {Serenelli}, Aldo and {Bautista}, Manuel and {Plez}, Bertrand and {Heiter}, Ulrike and {Gerber}, Jeffrey M. and {Ludwig}, Hans-G{\"u}nter and {Basu}, Sarbani and {Ferguson}, Jason W. and {Gallego}, Helena Carvajal and {Gamrath}, S{\'e}bastien and {Palmeri}, Patrick and {Quinet}, Pascal},
        title = "{Observational constraints on the origin of the elements. IV. Standard composition of the Sun}",
      journal = {\aap},
         year = 2022,
        month = may,
       volume = {661},
          eid = {A140},
        pages = {A140},
          doi = {10.1051/0004-6361/202142971},
archivePrefix = {arXiv},
       eprint = {2203.02255},
 primaryClass = {astro-ph.SR},
       adsurl = {https://ui.adsabs.harvard.edu/abs/2022A&A...661A.140M}
}

@ARTICLE{bastian_22,
       author = {{Bastian}, Nate and {Pfeffer}, Joel},
        title = "{Star cluster ecology: revisiting the origin of iron and age complex clusters}",
      journal = {\mnras},
         year = 2022,
        month = jan,
       volume = {509},
       number = {1},
        pages = {614-618},
          doi = {10.1093/mnras/stab3081},
archivePrefix = {arXiv},
       eprint = {2110.10616},
 primaryClass = {astro-ph.GA},
       adsurl = {https://ui.adsabs.harvard.edu/abs/2022MNRAS.509..614B}
}

@ARTICLE{crociati_23,
       author = {{Crociati}, Chiara and {Valenti}, Elena and {Ferraro}, Francesco R. and {Pallanca}, Cristina and {Lanzoni}, Barbara and {Cadelano}, Mario and {Fanelli}, Cristiano and {Origlia}, Livia and {Leanza}, Silvia and {Dalessandro}, Emanuele and {Mucciarelli}, Alessio and {Rich}, R. Michael},
        title = "{First Evidence of Multi-iron Subpopulations in the Bulge Fossil Fragment Candidate Liller 1}",
      journal = {\apj},
         year = 2023,
        month = jul,
       volume = {951},
       number = {1},
          eid = {17},
        pages = {17},
          doi = {10.3847/1538-4357/acd382},
archivePrefix = {arXiv},
       eprint = {2305.04595},
 primaryClass = {astro-ph.GA},
       adsurl = {https://ui.adsabs.harvard.edu/abs/2023ApJ...951...17C}
}

@ARTICLE{dalessandro_22,
       author = {{Dalessandro}, Emanuele and {Crociati}, Chiara and {Cignoni}, Michele and {Ferraro}, Francesco R. and {Lanzoni}, Barbara and {Origlia}, Livia and {Pallanca}, Cristina and {Rich}, R. Michael and {Saracino}, Sara and {Valenti}, Elena},
        title = "{Clues to the Formation of Liller 1 from Modeling Its Complex Star Formation History}",
      journal = {\apj},
         year = 2022,
        month = dec,
       volume = {940},
       number = {2},
          eid = {170},
        pages = {170},
          doi = {10.3847/1538-4357/ac9907},
archivePrefix = {arXiv},
       eprint = {2210.05694},
 primaryClass = {astro-ph.GA},
       adsurl = {https://ui.adsabs.harvard.edu/abs/2022ApJ...940..170D}
}

@ARTICLE{origlia_19,
       author = {{Origlia}, L. and {Mucciarelli}, A. and {Fiorentino}, G. and {Ferraro}, F.~R. and {Dalessandro}, E. and {Lanzoni}, B. and {Rich}, R.~M. and {Massari}, D. and {Contreras Ramos}, R. and {Matsunaga}, N.},
        title = "{Variable Stars in Terzan 5: Additional Evidence of Multi-age and Multi-iron Stellar Populations}",
      journal = {\apj},
         year = 2019,
        month = jan,
       volume = {871},
       number = {1},
          eid = {114},
        pages = {114},
          doi = {10.3847/1538-4357/aaf730},
       adsurl = {https://ui.adsabs.harvard.edu/abs/2019ApJ...871..114O}
}

@ARTICLE{pallanca_21,
       author = {{Pallanca}, Cristina and {Ferraro}, Francesco R. and {Lanzoni}, Barbara and {Crociati}, Chiara and {Saracino}, Sara and {Dalessandro}, Emanuele and {Origlia}, Livia and {Rich}, Michael R. and {Valenti}, Elena and {Geisler}, Douglas and {Mauro}, Francesco and {Villanova}, Sandro and {Moni Bidin}, Christian and {Beccari}, Giacomo},
        title = "{High-resolution Extinction Map in the Direction of the Strongly Obscured Bulge Fossil Fragment Liller 1}",
      journal = {\apj},
         year = 2021,
        month = aug,
       volume = {917},
       number = {2},
          eid = {92},
        pages = {92},
          doi = {10.3847/1538-4357/ac0889},
archivePrefix = {arXiv},
       eprint = {2106.02448},
 primaryClass = {astro-ph.GA},
       adsurl = {https://ui.adsabs.harvard.edu/abs/2021ApJ...917...92P}
}

@ARTICLE{ferraro_21,
       author = {{Ferraro}, F.~R. and {Pallanca}, C. and {Lanzoni}, B. and {Crociati}, C. and {Dalessandro}, E. and {Origlia}, L. and {Rich}, R.~M. and {Saracino}, S. and {Mucciarelli}, A. and {Valenti}, E. and {Geisler}, D. and {Mauro}, F. and {Villanova}, S. and {Moni Bidin}, C. and {Beccari}, G.},
        title = "{A new class of fossil fragments from the hierarchical assembly of the Galactic bulge}",
      journal = {Nature Astronomy},
         year = 2021,
        month = jan,
       volume = {5},
        pages = {311-318},
          doi = {10.1038/s41550-020-01267-y},
archivePrefix = {arXiv},
       eprint = {2011.09966},
 primaryClass = {astro-ph.GA},
       adsurl = {https://ui.adsabs.harvard.edu/abs/2021NatAs...5..311F}
}

@ARTICLE{ferraro_16,
       author = {{Ferraro}, F.~R. and {Massari}, D. and {Dalessandro}, E. and {Lanzoni}, B. and {Origlia}, L. and {Rich}, R.~M. and {Mucciarelli}, A.},
        title = "{The Age of the Young Bulge-like Population in the Stellar System Terzan 5: Linking the Galactic Bulge to the High-z Universe}",
      journal = {\apj},
         year = 2016,
        month = sep,
       volume = {828},
       number = {2},
          eid = {75},
        pages = {75},
          doi = {10.3847/0004-637X/828/2/75},
archivePrefix = {arXiv},
       eprint = {1609.01515},
 primaryClass = {astro-ph.GA},
       adsurl = {https://ui.adsabs.harvard.edu/abs/2016ApJ...828...75F}
}

@ARTICLE{immeli_04,
       author = {{Immeli}, A. and {Samland}, M. and {Gerhard}, O. and {Westera}, P.},
        title = "{Gas physics, disk fragmentation,  and bulge formation in young galaxies}",
      journal = {\aap},
         year = 2004,
        month = jan,
       volume = {413},
        pages = {547-561},
          doi = {10.1051/0004-6361:20034282},
archivePrefix = {arXiv},
       eprint = {astro-ph/0312139},
 primaryClass = {astro-ph},
       adsurl = {https://ui.adsabs.harvard.edu/abs/2004A&A...413..547I}
}

@ARTICLE{deimer24,
       author = {{Alvarez Garay}, D.~A. and {Fanelli}, C. and {Origlia}, L. and {Pallanca}, C. and {Mucciarelli}, A. and {Chiappino}, L. and {Crociati}, C. and {Lanzoni}, B. and {Ferraro}, F.~R. and {Rich}, R.~M. and {Dalessandro}, E.},
        title = "{X-shooter spectroscopy of Liller 1 giant stars}",
      journal = {\aap},
         year = 2024,
        month = jun,
       volume = {686},
          eid = {A198},
        pages = {A198},
          doi = {10.1051/0004-6361/202449595},
archivePrefix = {arXiv},
       eprint = {2404.14130},
 primaryClass = {astro-ph.GA},
       adsurl = {https://ui.adsabs.harvard.edu/abs/2024A&A...686A.198A}
}

@ARTICLE{1998A&A...330.1109A,
       author = {{Alvarez}, R. and {Plez}, B.},
        title = "{Near-infrared narrow-band photometry of M-giant and Mira stars: models meet observations}",
      journal = {\aap},
         year = 1998,
        month = feb,
       volume = {330},
        pages = {1109-1119},
          doi = {10.48550/arXiv.astro-ph/9710157},
archivePrefix = {arXiv},
       eprint = {astro-ph/9710157},
 primaryClass = {astro-ph},
       adsurl = {https://ui.adsabs.harvard.edu/abs/1998A&A...330.1109A}
}

@software{2012ascl.soft05004P,
       author = {{Plez}, B.},
        title = "{Turbospectrum: Code for spectral synthesis}",
 howpublished = {Astrophysics Source Code Library, record ascl:1205.004},
         year = 2012,
        month = may,
          eid = {ascl:1205.004},
archivePrefix = {ascl},
       eprint = {1205.004},
       adsurl = {https://ui.adsabs.harvard.edu/abs/2012ascl.soft05004P}
}

@ARTICLE{2015BaltA..24..453R,
       author = {{Ryabchikova}, T. and {Pakhomov}, Yu.},
        title = "{Archives of astronomical spectral observations and atomic / molecular databases for their analysis}",
      journal = {Baltic Astronomy},
         year = 2015,
        month = jan,
       volume = {24},
        pages = {453-461},
          doi = {10.1515/astro-2017-0249},
       adsurl = {https://ui.adsabs.harvard.edu/abs/2015BaltA..24..453R}
}

@ARTICLE{2008A&A...486..951G,
       author = {{Gustafsson}, B. and {Edvardsson}, B. and {Eriksson}, K. and {J{\o}rgensen}, U.~G. and {Nordlund}, {\r{A}}. and {Plez}, B.},
        title = "{A grid of MARCS model atmospheres for late-type stars. I. Methods and general properties}",
      journal = {\aap},
         year = 2008,
        month = aug,
       volume = {486},
       number = {3},
        pages = {951-970},
          doi = {10.1051/0004-6361:200809724},
archivePrefix = {arXiv},
       eprint = {0805.0554},
 primaryClass = {astro-ph},
       adsurl = {https://ui.adsabs.harvard.edu/abs/2008A&A...486..951G}
}

@ARTICLE{Thorsbro18,
       author = {{Thorsbro}, B. and {Ryde}, N. and {Schultheis}, M. and {Hartman}, H. and {Rich}, R.~M. and {Lomaeva}, M. and {Origlia}, L. and {J{\"o}nsson}, H.},
        title = "{Evidence against Anomalous Compositions for Giants in the Galactic Nuclear Star Cluster}",
      journal = {\apj},
         year = 2018,
        month = oct,
       volume = {866},
       number = {1},
          eid = {52},
        pages = {52},
          doi = {10.3847/1538-4357/aadb97},
archivePrefix = {arXiv},
       eprint = {1808.07489},
 primaryClass = {astro-ph.GA},
       adsurl = {https://ui.adsabs.harvard.edu/abs/2018ApJ...866...52T}
}

@ARTICLE{Thorsbro20_1,
       author = {{Thorsbro}, Brian},
        title = "{Atomic Data Needs in Astrophysics: The Galactic Center ``Scandium Mystery''}",
      journal = {Atoms},
         year = 2020,
        month = jan,
       volume = {8},
       number = {1},
        pages = {4},
          doi = {10.3390/atoms8010004},
       adsurl = {https://ui.adsabs.harvard.edu/abs/2020Atoms...8....4T}
}

@ARTICLE{Thorsbro20_2,
       author = {{Thorsbro}, B. and {Ryde}, N. and {Rich}, R.~M. and {Schultheis}, M. and {Renaud}, F. and {Spitoni}, E. and {Fritz}, T.~K. and {Mastrobuono-Battisti}, A. and {Origlia}, L. and {Matteucci}, F. and {Sch{\"o}del}, R.},
        title = "{Detailed Abundances in the Galactic Center: Evidence of a Metal-rich Alpha-enhanced Stellar Population}",
      journal = {\apj},
         year = 2020,
        month = may,
       volume = {894},
       number = {1},
          eid = {26},
        pages = {26},
          doi = {10.3847/1538-4357/ab8226},
archivePrefix = {arXiv},
       eprint = {2003.11085},
 primaryClass = {astro-ph.GA},
       adsurl = {https://ui.adsabs.harvard.edu/abs/2020ApJ...894...26T}
}

@ARTICLE{cha95,
       author = {{Charbonnel}, C.},
        title = "{A Consistent Explanation for 12C/ 13C, 7Li and 3He Anomalies in Red Giant Stars}",
      journal = {\apjl},
         year = 1995,
        month = nov,
       volume = {453},
        pages = {L41},
          doi = {10.1086/309744},
archivePrefix = {arXiv},
       eprint = {astro-ph/9511080},
 primaryClass = {astro-ph},
       adsurl = {https://ui.adsabs.harvard.edu/abs/1995ApJ...453L..41C}
}

@ARTICLE{den96,
       author = {{Denissenkov}, P.~A. and {Weiss}, A.},
        title = "{Deep diffusive mixing in globular-cluster red giants.}",
      journal = {\aap},
         year = 1996,
        month = apr,
       volume = {308},
        pages = {773-784},
       adsurl = {https://ui.adsabs.harvard.edu/abs/1996A&A...308..773D}
}

@ARTICLE{cav98,
       author = {{Cavallo}, Robert M. and {Sweigart}, Allen V. and {Bell}, Roger A.},
        title = "{Proton-Capture Nucleosynthesis in Globular Cluster Red Giant Stars}",
      journal = {\apj},
         year = 1998,
        month = jan,
       volume = {492},
       number = {2},
        pages = {575-595},
          doi = {10.1086/305053},
archivePrefix = {arXiv},
       eprint = {astro-ph/9708099},
 primaryClass = {astro-ph},
       adsurl = {https://ui.adsabs.harvard.edu/abs/1998ApJ...492..575C}
}

@ARTICLE{boo99,
       author = {{Boothroyd}, Arnold I. and {Sackmann}, I.-Juliana},
        title = "{The CNO Isotopes: Deep Circulation in Red Giants and First and Second Dredge-up}",
      journal = {\apj},
         year = 1999,
        month = jan,
       volume = {510},
       number = {1},
        pages = {232-250},
          doi = {10.1086/306546},
       adsurl = {https://ui.adsabs.harvard.edu/abs/1999ApJ...510..232B}
}

@ARTICLE{Bist14,
       author = {{Bisterzo}, S. and {Travaglio}, C. and {Gallino}, R. and {Wiescher}, M. and {K{\"a}ppeler}, F.},
        title = "{Galactic Chemical Evolution and Solar s-process Abundances: Dependence on the $^{13}$C-pocket Structure}",
      journal = {\apj},
         year = 2014,
        month = may,
       volume = {787},
       number = {1},
          eid = {10},
        pages = {10},
          doi = {10.1088/0004-637X/787/1/10},
archivePrefix = {arXiv},
       eprint = {1403.1764},
 primaryClass = {astro-ph.SR},
       adsurl = {https://ui.adsabs.harvard.edu/abs/2014ApJ...787...10B}
}

@ARTICLE{Prant20,
       author = {{Prantzos}, N. and {Abia}, C. and {Cristallo}, S. and {Limongi}, M. and {Chieffi}, A.},
        title = "{Chemical evolution with rotating massive star yields II. A new assessment of the solar s- and r-process components}",
      journal = {\mnras},
         year = 2020,
        month = jan,
       volume = {491},
       number = {2},
        pages = {1832-1850},
          doi = {10.1093/mnras/stz3154},
archivePrefix = {arXiv},
       eprint = {1911.02545},
 primaryClass = {astro-ph.GA},
       adsurl = {https://ui.adsabs.harvard.edu/abs/2020MNRAS.491.1832P}
}

@ARTICLE{barbuy_15,
       author = {{Barbuy}, B. and {Fria{\c{c}}a}, A.~C.~S. and {da Silveira}, C.~R. and {Hill}, V. and {Zoccali}, M. and {Minniti}, D. and {Renzini}, A. and {Ortolani}, S. and {G{\'o}mez}, A.},
        title = "{Zinc abundances in Galactic bulge field red giants: Implications for damped Lyman-{\ensuremath{\alpha}} systems}",
      journal = {\aap},
         year = 2015,
        month = aug,
       volume = {580},
          eid = {A40},
        pages = {A40},
          doi = {10.1051/0004-6361/201525694},
archivePrefix = {arXiv},
       eprint = {1506.01612},
 primaryClass = {astro-ph.GA},
       adsurl = {https://ui.adsabs.harvard.edu/abs/2015A&A...580A..40B}
}

@ARTICLE{beh16,
       author = {{Behrendt}, M. and {Burkert}, A. and {Schartmann}, M.},
        title = "{Clusters of Small Clumps Can Explain the Peculiar Properties of Giant Clumps in High-redshift Galaxies}",
      journal = {\apjl},
         year = 2016,
        month = mar,
       volume = {819},
       number = {1},
          eid = {L2},
        pages = {L2},
          doi = {10.3847/2041-8205/819/1/L2},
archivePrefix = {arXiv},
       eprint = {1512.03430},
 primaryClass = {astro-ph.GA},
       adsurl = {https://ui.adsabs.harvard.edu/abs/2016ApJ...819L...2B}
}

@ARTICLE{dek09,
       author = {{Dekel}, Avishai and {Sari}, Re'em and {Ceverino}, Daniel},
        title = "{Formation of Massive Galaxies at High Redshift: Cold Streams, Clumpy Disks, and Compact Spheroids}",
      journal = {\apj},
         year = 2009,
        month = sep,
       volume = {703},
       number = {1},
        pages = {785-801},
          doi = {10.1088/0004-637X/703/1/785},
archivePrefix = {arXiv},
       eprint = {0901.2458},
 primaryClass = {astro-ph.GA},
       adsurl = {https://ui.adsabs.harvard.edu/abs/2009ApJ...703..785D}
}

@ARTICLE{ger12,
       author = {{Gerhard}, Ortwin and {Martinez-Valpuesta}, Inma},
        title = "{The Inner Galactic Bulge: Evidence for a Nuclear Bar?}",
      journal = {\apjl},
         year = 2012,
        month = jan,
       volume = {744},
       number = {1},
          eid = {L8},
        pages = {L8},
          doi = {10.1088/2041-8205/744/1/L8},
archivePrefix = {arXiv},
       eprint = {1112.0179},
 primaryClass = {astro-ph.GA},
       adsurl = {https://ui.adsabs.harvard.edu/abs/2012ApJ...744L...8G}
}

@ARTICLE{saha13,
       author = {{Saha}, Kanak and {Gerhard}, Ortwin},
        title = "{Secular evolution and cylindrical rotation in boxy/peanut bulges: impact of initially rotating classical bulges}",
      journal = {\mnras},
         year = 2013,
        month = apr,
       volume = {430},
       number = {3},
        pages = {2039-2046},
          doi = {10.1093/mnras/stt029},
archivePrefix = {arXiv},
       eprint = {1212.2317},
 primaryClass = {astro-ph.CO},
       adsurl = {https://ui.adsabs.harvard.edu/abs/2013MNRAS.430.2039S}
}

@ARTICLE{kalita22,
       author = {{Kalita}, Boris S. and {Daddi}, Emanuele and {Bournaud}, Frederic and {Rich}, Robert Michael and {Valentino}, Francesco and {G{\'o}mez-Guijarro}, Carlos and {Codis}, Sandrine and {Delvecchio}, Ivan and {Elbaz}, David and {Strazzullo}, Veronica and {de Souza Magalhaes}, Victor and {Pety}, J{\'e}r{\^o}me and {Tan}, Qinghua},
        title = "{Bulge formation inside quiescent lopsided stellar disks: Connecting accretion, star formation, and morphological transformation in a z {\ensuremath{\sim}} 3 galaxy group}",
      journal = {\aap},
         year = 2022,
        month = oct,
       volume = {666},
          eid = {A44},
        pages = {A44},
          doi = {10.1051/0004-6361/202243100},
archivePrefix = {arXiv},
       eprint = {2206.05217},
 primaryClass = {astro-ph.GA},
       adsurl = {https://ui.adsabs.harvard.edu/abs/2022A&A...666A..44K}
}

@ARTICLE{tan24,
       author = {{Tan}, Qing-Hua and {Daddi}, Emanuele and {Magnelli}, Benjamin and {Correa}, Camila A. and {Bournaud}, Fr{\'e}d{\'e}ric and {Adscheid}, Sylvia and {Zhang}, Shao-Bo and {Elbaz}, David and {G{\'o}mez-Guijarro}, Carlos and {Kalita}, Boris S. and {Liu}, Daizhong and {Liu}, Zhaoxuan and {Pety}, J{\'e}r{\^o}me and {Puglisi}, Annagrazia and {Schinnerer}, Eva and {Silverman}, John D. and {Valentino}, Francesco},
        title = "{In situ spheroid formation in distant submillimetre-bright galaxies}",
      journal = {\nat},
         year = 2024,
        month = dec,
       volume = {636},
       number = {8041},
        pages = {69-74},
          doi = {10.1038/s41586-024-08201-6},
archivePrefix = {arXiv},
       eprint = {2407.16578},
 primaryClass = {astro-ph.GA},
       adsurl = {https://ui.adsabs.harvard.edu/abs/2024Natur.636...69T}
}
\end{document}